\documentclass[a4paper,11pt]{article}

\usepackage[margin=2.5cm]{geometry}

\usepackage{graphicx}
\usepackage{physics}
\usepackage{aas_macros}

\usepackage{subcaption}
\usepackage{amsmath,amssymb}
\usepackage{rotating}

\usepackage{authblk}
\usepackage{cprotect}

\usepackage{hyperref}

\usepackage[dvipsnames,table]{xcolor}

\hypersetup{
  colorlinks,
  citecolor=blue,
  linkcolor=red,
  urlcolor=olive}

\usepackage[title]{appendix}

\newcommand{\sevem}{\texttt{SEVEM}}

\newcommand{\planck}{\textit{Planck}}
\newcommand{\nside}{$N_{\rm side}$}

\newcommand{\healpix}{\texttt{HEALPix}}
\newcommand{\Scipy}{\texttt{Scipy}}

\usepackage{aas_macros,physics}

\title{Examination of frequency and scale dependence of CMB hemispherical power asymmetry}
\author[1]{Sanjeev Sanyal\thanks{sanjeevsanyal.rs.phy19@itbhu.ac.in}}
\author[1]{Pavan K. Aluri\thanks{pavanaluri.phy@itbhu.ac.in}}
\author[2,3]{Arman Shafieloo\thanks{shafieloo@kasi.re.kr}}
\affil[1]{Dept. of Physics, Indian Institute of Technology (BHU), Varanasi - 221005, India}
\affil[2]{Korea Astronomy and Space Science Institute (KASI), Yuseong-gu, 776 Daedeok daero, Daejeon 34055, Republic of Korea}
\affil[3]{University of Science and Technology (UST), Yuseong-gu 217 Gajeong-ro, Daejeon 34113, Republic of Korea}

\date{\today}

\begin{document}

\maketitle

\begin{abstract}
In this study, we revisit the well-known cosmic microwave background (CMB) anomaly referred to as \textit{Hemispherical Power Asymmetry} (HPA), using the CMB temperature maps from the \planck\ mission’s public release 4 (PR4) and WMAP's 9yr data release. Employing the Local Variance Estimator (LVE) method, we systematically reexamine the properties of HPA for any potential frequency dependence and also scale dependence in its amplitude and direction, modeling the HPA as a scale dependent dipole modulation following a power-law form rather than a scale invariant case.
Our analysis incorporates \emph{seven} cleaned frequency-specific CMB temperature maps from both the \planck\ and WMAP missions to test the robustness of the observed asymmetry across instruments and frequency channels. We find that the dipolar modulation characteristic of HPA is present in all cases examined, with consistent estimates of direction and variation in dipole amplitudes over scales. These results support the conclusion that the observed asymmetry is unlikely to be a product of instrumental artifacts or data processing, but rather a persistent, large-scale feature of the CMB sky with cosmological origin.
\end{abstract}

\section{Introduction}
\label{sec:intro}
Steady technological advances over the past few years have enabled the cosmologists to observe the cosmos across the entire electromagnetic spectrum with better precision. The cosmic microwave background (CMB) is an important signal from the early universe, with which cosmologists have been performing stringent tests since the first observations of CMB anisotropies made with the COBE satellite~\cite{Smoot1991}. It is a crucial source of data for studying the evolution of our Universe. CMB radiation is the fingerprint of the universe when it was very young ($\sim$ 3,80,000 years after Big Bang), and is found to be very uniform across the sky.
The fundamental assumption of modern cosmology, namely the Cosmological Principle ($CP$), is a statement about the nature of space viz., the universe is spatially isotropic and homogeneous when viewed beyond length scales of the typical size of a supercluster of galaxies ($\sim200$ Mpc).

Several studies investigated the validity of $CP$, and the results hinted at a breakdown of isotropy, i.e., there are indications for some preferred direction(s) for our Universe as opposed to the expectation based on the assumption of spatial isotropy via $CP$.
These signatures of isotropy violation persisted in CMB data from WMAP and \planck\ missions when examined using different statistical techniques. See, for example, Refs.~\cite{schwarz2016,bull2016,abdalla2022,cp2023} for a review of various anomalies seen in CMB sky that violate isotropy (along with other tensions of observational cosmology that challenge the current standard model).

Wilkinson Microwave Anisotropy Probe (WMAP) was the second space based mission after COBE to measure CMB anisotropies with much higher precision~\cite{Bennett_2003}. WMAP observed the CMB sky in five different frequency channels centered at 23~GHz (K-band), 33~GHz (Ka-band), 41~GHz (Q-band), 61~GHz (V-band), and 94~GHz (W-band) respectively. The local power analysis performed on the first year WMAP full sky temperature data~\cite{Eriksen_2004,Hansen_2004} revealed an excess power in the CMB sky in one hemisphere compared to the other. The same feature was also found in later WMAP data releases that has come to be called Hemispherical Power Asymmetry (HPA)~\cite{Eriksen_2007,HansonLewis2009,wmap7yranom}. This asymmetry was found to be a dipole of amplitude $A \approx 0.07$ along the direction $\sim(224^\circ,-22^\circ)$ in galactic coordinates~\cite{Hoftuft_2009_lowL}.

The \planck\ satellite, with more sensitivity and higher resolution compared to WMAP, observed the CMB sky across a broader range of frequency channels than WMAP, covering complementary and higher-frequency bands~\cite{plk13overview}. The satellite operated in nine frequency bands ranging from 30 GHz to 857 GHz. Three at lower frequencies - 30, 44, and 70 GHz comprise the Low Frequency Instrument (LFI), and six at higher frequencies - 100, 143, 217, 353, 545, and 857 GHz make up the High Frequency Instrument (HFI). \planck\ data also confirmed the presence of HPA in its temperature anisotropy measurements of CMB sky~\cite{plk2013isostat,plk2015isostat,plk2018isostat}.

Many methods were proposed to assess this power asymmetry seen in the CMB sky i.e., to characterize its robustness in greater detail and in alternate ways~\cite{Hajian2005,Lew2008,Bernui2008,Paci2010,Pranati2013,Akrami2014,Aiola_2015,Ghosh_2016,Tarun2019anom,rajib2023hpaai,Gimeno_Amo_2023}. In Ref.~\cite{Akrami2014}, a local variance estimator (LVE) method was proposed to probe HPA based on the expectation that the properties of the CMB sky should be statistically same irrespective of the location and shape of the CMB patch observed/considered. This expectation follows directly from $CP$. From the observed CMB sky, a map of locally computed variances was derived for different choices of circular disc sizes uniformly covering the entire sky to probe HPA by angular scale.
Assuming HPA to be a pure dipolar feature, a dipole was fit to the data derived local variance map, from which the amplitude of variance asymmetry was found at a more than $3\sigma$ significance~\cite{Akrami2014,Adhikari2015,plk2015isostat,plk2018isostat} when compared with simulations based on $\Lambda$CDM model.

The observed modulation signal underlying the CMB sky was found to be present at low multipoles of $l \lesssim 64$, that correspond to large angular scales of the CMB sky~\cite{Hoftuft_2009_lowL,HansonLewis2009}. The signature of HPA was also found to be present at small angular scales by way of anomalous alignment of dipole directions recovered from binned power asymmetry map obtained using higher multipole ranges~\cite{plk2013isostat,plk2015isostat,Hansen_2009,Flender_2013,QuartinNotari2015}.
Variation in dipole amplitude of HPA with multipole range considered indicates a scale dependence of the same. Therefore, this hemispherical power asymmetry is not fixed in strength across angular scales. It is more striking on large scales and lessens on going to smaller scales. However, few have tested/modeled HPA as scale dependent phenomena with explicit analytic form (viz., an inverse power-law)~\cite{Aiola_2015,Tarun2019anom,scal_dep_dm_l23_2019}.

Likewise, few have investigated the frequency-specific \emph{cleaned} CMB maps (apart from the usual internal liner combination (ILC) method cleaned maps - in real or harmonic space) in view of HPA anomaly. WMAP's foreground-reduced Q, V and W band maps from WMAP's 3yr and 5yr data release were studied, for example, in Ref.~\cite{Eriksen_2007,HansonLewis2009,Hoftuft_2009_lowL,Bernui2008,Hansen_2009}.
\planck\ foreground cleaned frequency-specific CMB temperature maps (two or three of the \sevem\ 100, 143 and 217~GHz CMB maps) were also studied to test for frequency independence of the observed hemispherical power asymmetry as a result of foreground residuals~\cite{Adhikari2015,plk2015isostat}.

Here, our objective is to test for frequency and scale (multipole) (in)dependence of the CMB hemispherical power asymmetry anomaly. To that end, we use ``seven'' cleaned frequency-specific CMB maps viz., WMAP's 9yr 41~GHz (Q-band), 61~GHz (V-band), and 94~GHz (W-band) foreground-reduced CMB maps, and the \sevem\ cleaned 70~GHz, 100~GHz, 143~GHz, and 217~GHz frequency-specific CMB maps from \planck's public release 4 (PR4). Rest of the paper is organized as follows. In section~\ref{sec:methods}, the LVE method for probing HPA is described, followed by a description of data and complementary simulations used in the current analysis in section~\ref{sec:data-sim}. Then, our results are presented in section~\ref{sec:results}, and finally conclusions thus drawn are summarized in section~\ref{sec:concl}.

\section{Methodology}
\label{sec:methods}
The hemispherical power asymmetry seen in the observed CMB sky is phenomenologically modeled as a modulation of an otherwise isotropic CMB sky~\cite{Gordon2005}, in the form of a pure dipole-like field that is mathematically expressed as
\begin{equation}
\Delta T_{\rm mod}(\hat{n}) = (1 +  \vec{d} \cdot \hat{n}) \Delta T_{\rm iso}(\hat{n}) = (1 +  A \, \hat{\lambda} \cdot \hat{n}) \Delta T_{\rm iso}(\hat{n})\,,
\label{eq:mod}
\end{equation}
where $\Delta T_{\rm mod}(\hat{n})$ and $\Delta T_{\rm iso}(\hat{n})$ are the temperature fluctuations of the modulated (observed) and unmodulated (isotropic) CMB sky in the `$\hat{n}$' direction, respectively. The term $\vec{d} = A \, \hat{\lambda}$ is the dipole modulation field underlying the CMB sky with $A$, $\hat{\lambda}$ denoting its magnitude and direction, respectively. 

One way to probe the power asymmetry is to check the consistency of a chosen statistic across the CMB map. We choose the statistic to be locally computed \emph{variance} of the CMB sky. Then $CP$ suggests that the variances thus estimated from patches of various size uniformly covering the entire sky, should be statistically consistent with each other. Thus we use the Local Variance Estimator (abbreviated as LVE) to probe this power asymmetry in the CMB temperature sky. LVE exploits the information locally from an input map in real (pixel) space by considering circular discs of radius `$r$', to estimate the statistical quantities of interest viz., the variance of CMB sky with radii ranging from $r=1^\circ$ to 40$^\circ$. Assuming the presence of a dipolar feature inducing a small deviation from isotropy in the data, one can expect the local variance maps (LVE maps/LVMs) computed for different choices of disc sizes to be consistent with a similar dipolar feature.

CMB temperature maps from WMAP and \planck\ were digitized using the \healpix\ pixelization scheme i.e., $\Delta T(\hat{n})\equiv \Delta T(p)$ where `$p$' is the pixel index corresponding to the position $\hat{n}$ in the sky. Local variances are computed from all the pixels of an input CMB map that fall inside a circular patch of radius `$r$' defined at some location $\hat{N}$. One can then express the local variance estimator as
\begin{equation}
{\sigma_{r}}^{2}{(\hat{N})} =\frac{1}{N_p} \sum_{p{\in}r@{\hat{N}}}(T(p) - {\bar{T_{r}}})^{2}\,,
\label{eq:var}
\end{equation}
where ${\bar{T_{r}}}$ is the mean value of the temperature fluctuations calculated from the pixels within the circular disc of radius `$r$', `$p$' denotes the pixel indices of the observed CMB sky, and $N_{p}$ stands for the total number of pixels within the same disc centered in the direction $\hat{N}$ of the local variance map. We shall use CMB sky maps only after masking to remove potentially contaminated regions. As a result of masking, pixels mostly near the Galactic region will not be included in the analysis. Further, when using smaller disc sizes, it is often the case that a large fraction of the pixels within a given disc are masked especially near the Galactic plane. For a reliable statistical estimate, only circular discs in which at least 100 unmasked pixels or 50\% pixels are available compared to those in a full circular disc, whichever number is greater. Note that this is a more conservative choice of pixel fraction in comparison to the 10\% criterion used originally~\cite{Akrami2014}.

Substituting Eq.~(\ref{eq:mod}) in Eq.~(\ref{eq:var}) with the assumption that the dipole amplitude is small and keeping terms only up to linear order in `$A$', we will get an approximate expression for variances estimated locally from a dipole-modulated map as 
\begin{equation}
\sigma^{2}_{{\rm obs},r}{(\hat{N})} \approx \sigma^{2}_{{\rm iso},r} (\hat{N})(1 + 2 A \, \hat{\lambda} \cdot \hat{N})\,.   
\label{eq:dip-var}
\end{equation}
The above equation is valid only up to a maximum radius `$r$' of the circular disc contingent up on our tolerance limit for recovering the dipole signal underlying the CMB sky~\cite{sanyal2024hpa}.

We then compute the dipole amplitude and direction from the normalized variance map that is defined as
\begin{equation}
\xi_{r}{(\hat{N})} = \frac{\sigma_{\rm obs,r}^2(\hat{N})- \langle \sigma_{\rm iso,r}^2(\hat{N}) \rangle}{\langle \sigma^2_{\rm iso,r}{(\hat{N})} \rangle}\,.  
\label{eq:norm-var-map}
\end{equation}
The subscript `$r$' in $\xi_{r}{(\hat{N})}$ denotes that this quantity represents a normalized variance map corresponding to that specific disc size. The bias correction term, $\langle{\sigma}^2_{\rm iso,r}(\hat{N}) \rangle$ is the pixel-by-pixel average calculated from isotropic (unmodulated) simulations to correct for the expected level of random fluctuation in any given realization of CMB sky, instrumental noise etc. From the normalized local variance maps, we extract the dipolar field, i.e, its amplitude and direction using \healpix's \verb+remove_dipole+ routine. From Eq.~(\ref{eq:dip-var}) and (\ref{eq:norm-var-map}), we find that $\xi_{r}{(\hat{N})} \equiv 2A\hat{\lambda}\cdot\hat{n}$. Thus, the dipole amplitude of normalized local variance maps, $\xi_{r}{(\hat{N})}$, is $A_{\rm LV}=2A$ whose direction is same as the HPA underlying the CMB sky.

\begingroup

\renewcommand{\arraystretch}{1.25} 
\begin{table}[t]
\centering
\begin{tabular}{| r || c | c | c | c | c |}
\hline
\nside\ & 32 & 16 & 8 & 4 & 2 \\

\hline
Disc Size ($r^\circ$) & 1 & 2 & 4, 6 & 8, 10, 12, 15 & 20, 25, 30, 40 \\
\hline
\end{tabular}
\caption{\healpix\ grid resolution (\nside) chosen for deriving local variance maps for different choices of circular disc sizes ($r$) considered in this work.}
\label{tab:hpx-pixsize}
\end{table}

\endgroup

Recently in Ref.~\cite{sanyal2024hpa}, we appraised the LVE method extensively regarding its various procedural aspects such as choice of disc radius `$r$' and the appropriate \healpix\ grid to use for the corresponding local variance map being computed, use of LVM covariance matrix in estimation of underlying dipole (amplitude and direction), any limitation of the method, etc. 
In the process of dipole estimation from local variance maps, we use the diagonal elements of the covariance matrix as weights to perform inverse variance weighting.
Further in our previous work~\cite{sanyal2024hpa}, we identified a range of disc sizes over which the LVE method is reliable (if it were $A\sim0.07$). Accordingly, in the present study, we restricted our evaluation of local variance maps up to $r=40^\circ$. Table~{\ref{tab:hpx-pixsize}} lists the different disc radii used in the present analysis and the corresponding \healpix\ grid resolution at which we will map local variances computed from the input CMB maps.
The dipole amplitudes thus obtained from normalized local variance maps are biased. They are corrected for the expected random dipole that can arise in a specific realization of the CMB sky from isotropic simulations. This correction has to be performed at the power level ie.,
\begin{equation}
A_{\rm LV,corr}=\sqrt{A_{\rm LV,obs}^2 - \langle A_{\rm LV,iso}^2 \rangle} \,,
\label{eq:alv-bias-corr}
\end{equation}
and not at map/amplitude level ($A_{\rm LV,corr}\neq A_{\rm LV,obs} - {\langle A_{\rm LV,iso} \rangle}$)~\cite{sanyal2024hpa}.

Power asymmetry in CMB sky is explored in real space as well as in harmonic space. Most of the analysis viewed the HPA feature as a dipole modulation of an otherwise isotropic CMB sky with scale independent amplitude (at low multipoles or at large angular scales). Few searched for departures from scale invariance.
To test the scale dependence of HPA amplitude, we fit the bias corrected dipole amplitude versus multipole with an inverse power-law function as~\cite{Aiola_2015,Tarun2019anom}
\begin{equation}
    A_{\rm LV, corr} \equiv A(\ell) = A_0 \left( \frac{\ell_0}{\ell} \right)^{n}\,,
\label{eq:pl}
\end{equation}
where `$A_{0}$' is the amplitude at a pivot/reference multipole `$\ell_0$', and `$\ell$' is the scale/multipole that is related to the disc radius `$r$' as $\ell \sim 180^\circ / r$. The pivot scale for this power-law fit is chosen to be $\ell_0=10$ in our present analysis. A different choice of $\ell_0$ can always be taken care of as an overall multiplicative factor. According to the power-law model of Eq.~(\ref{eq:pl}), we have two free parameters to fit viz., the dipole amplitude `$A_{0}$' (at the pivot scale `$\ell_0$') and the index `$n$'. The dipole direction is taken as is obtained from the LVE maps.

\section{Data, Simulations and Masks}
\label{sec:data-sim}

\begin{figure}
\centering
\includegraphics[width=0.48\textwidth]{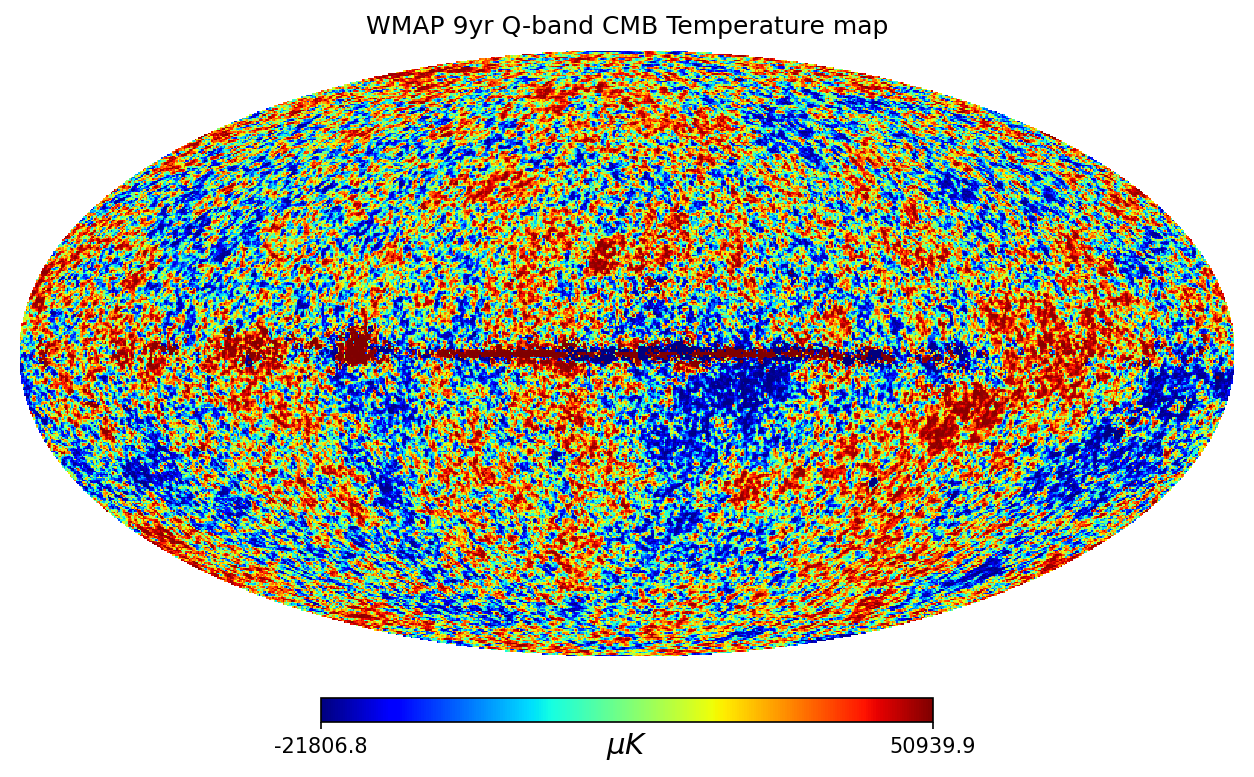}
~
\includegraphics[width=0.48\textwidth]{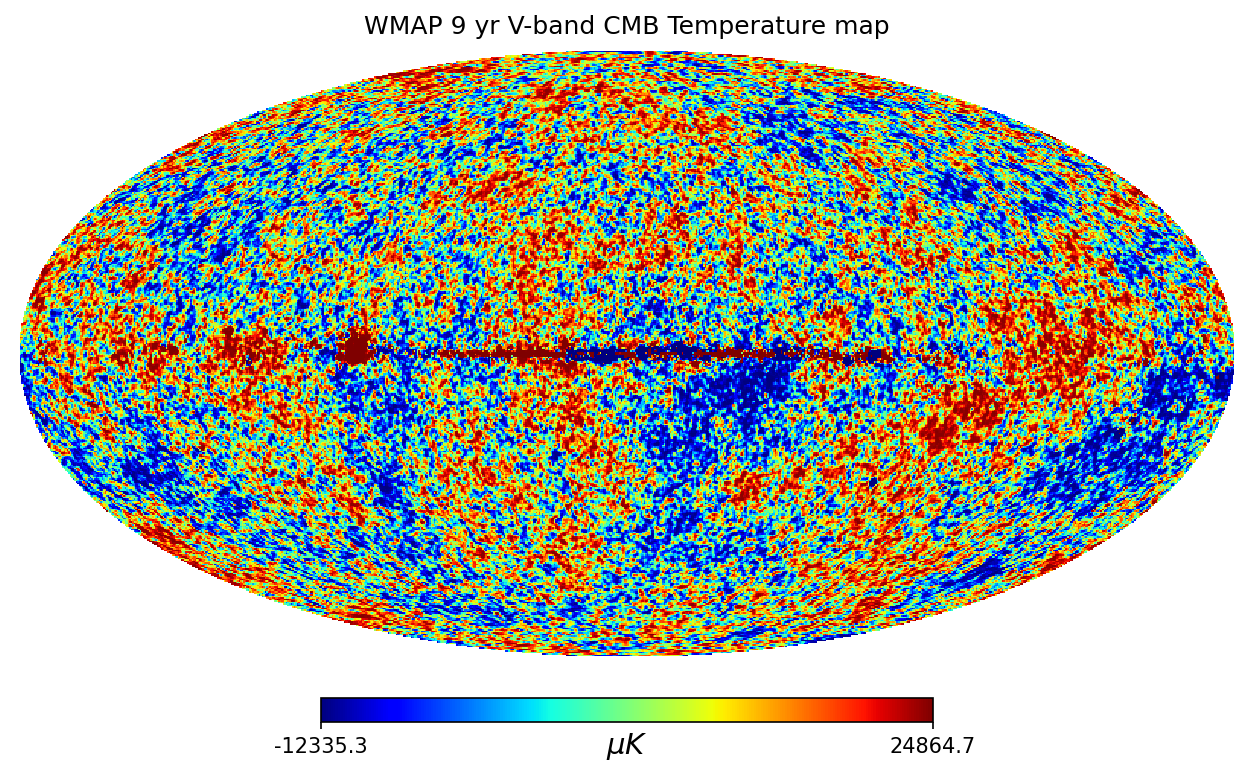}
~
\includegraphics[width=0.48\textwidth]{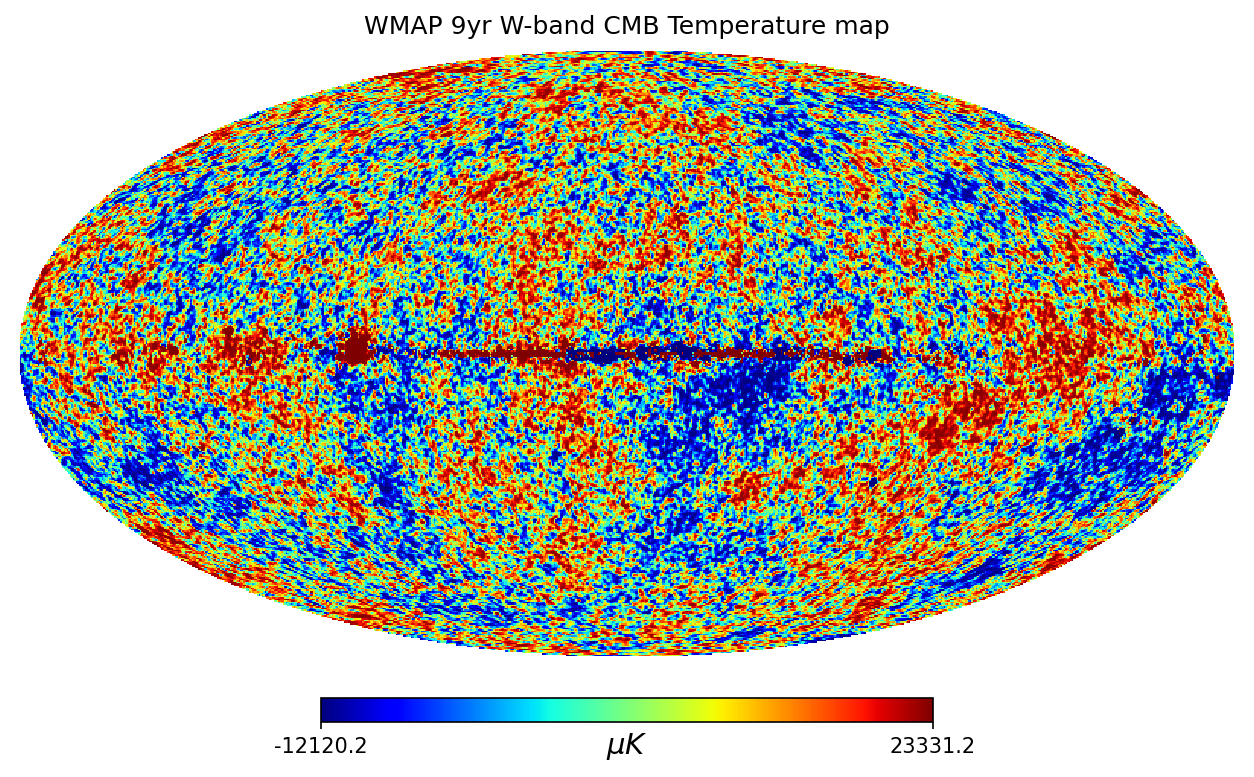}
~
\includegraphics[width=0.48\textwidth]{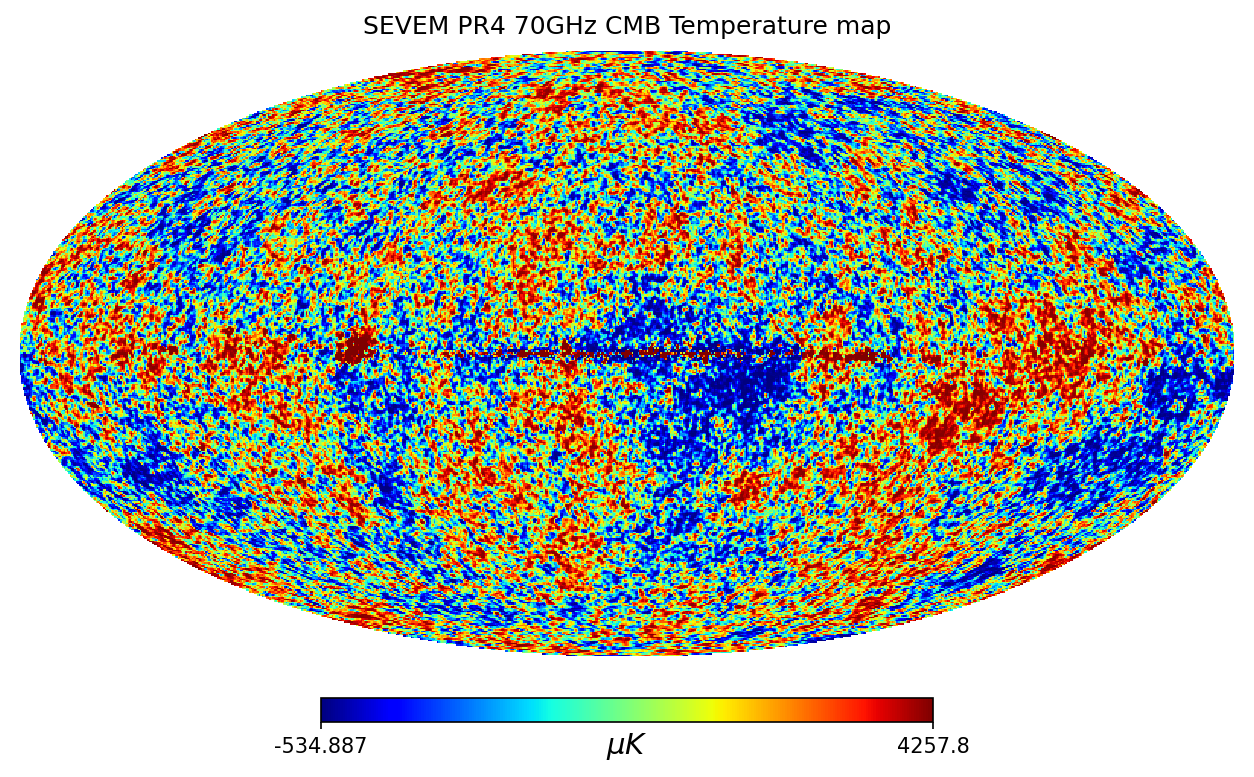}
~
\includegraphics[width=0.48\textwidth]{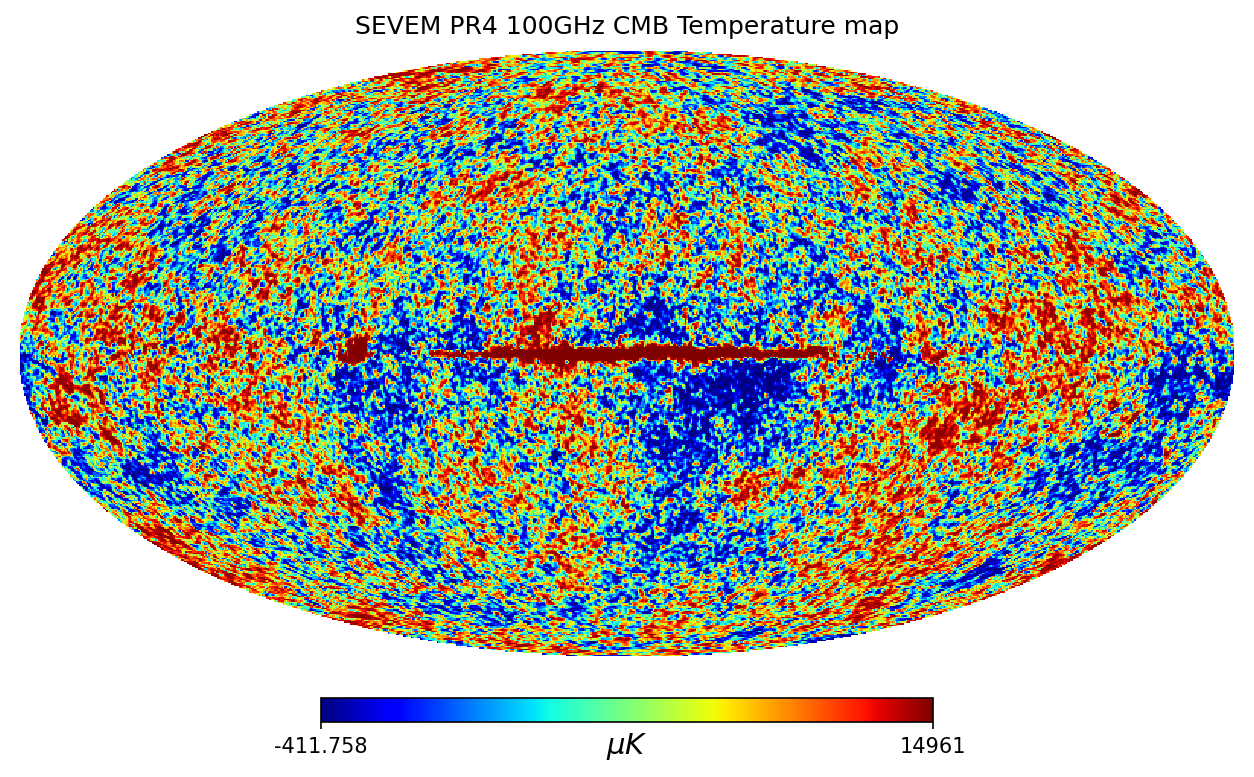}
~
\includegraphics[width=0.48\textwidth]{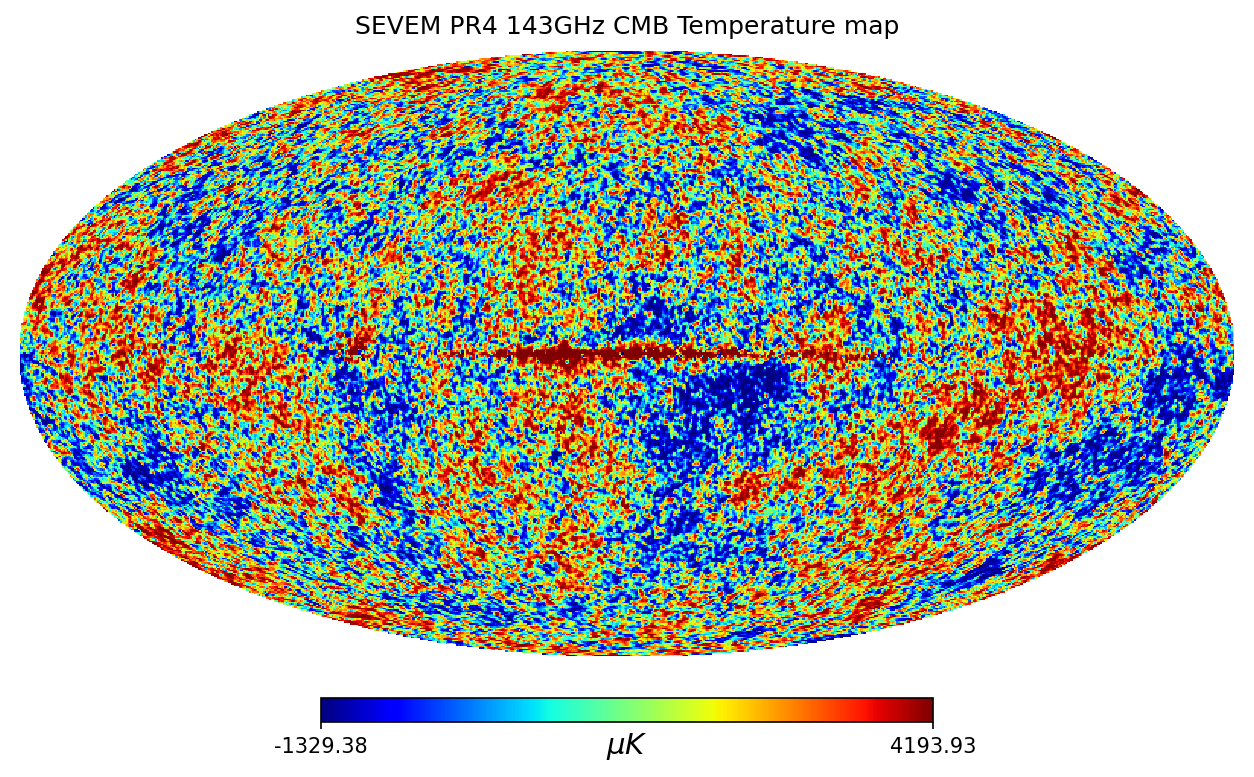}
~
\includegraphics[width=0.48\textwidth]{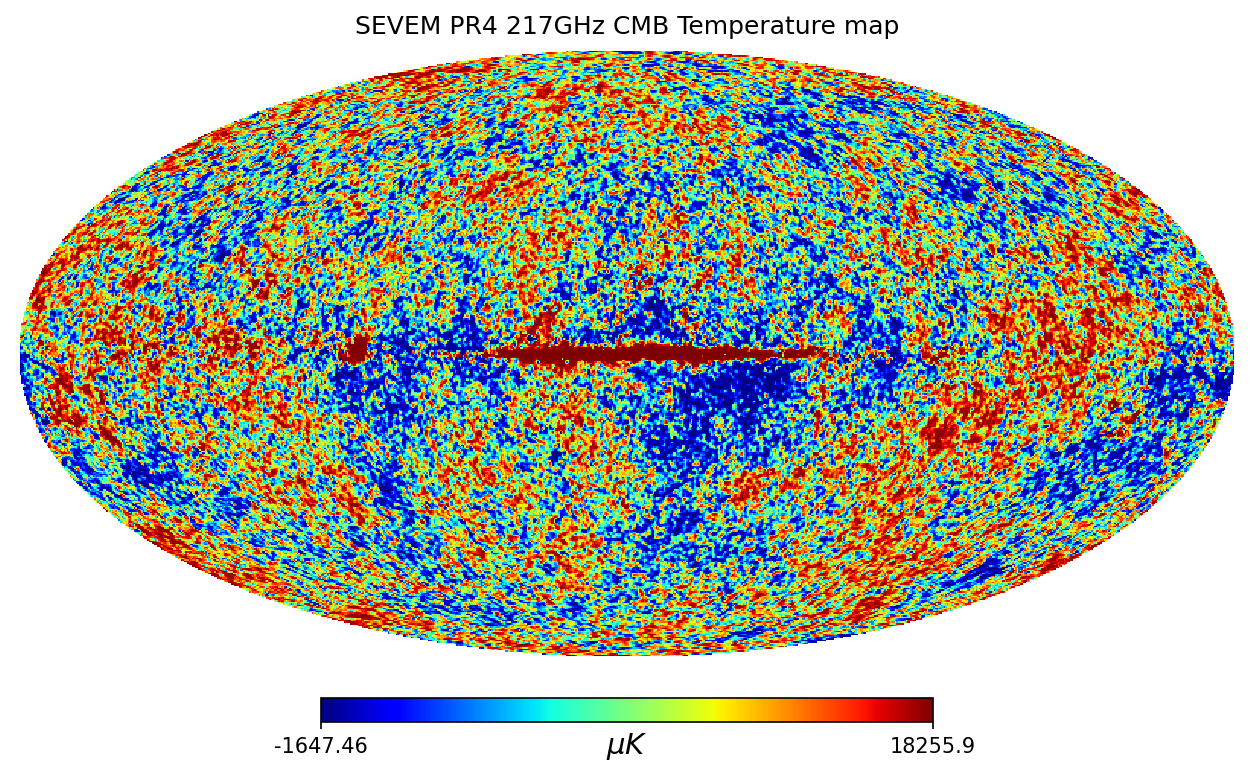}
\caption{Frequency-specific cleaned CMB maps from observations, all of which are synthesized at \healpix\ \nside=512 with a Gaussian beam of FWHM$=30'$. Sequentially (top to bottom \& left to right), the WMAP 9yr Q (41~GHz), V (61~GHz) and W (94~GHz) band maps are shown first, followed by the \planck\ PR4 \sevem\ cleaned 70, 100, 143 and 217~GHz maps are displayed.}
\label{fig:datamaps}
\end{figure}

\subsection{Data}
\label{sec:data}
To probe frequency dependence of HPA, if any, we analyze \emph{seven} different frequency-specific cleaned CMB maps from WMAP and \planck\ missions.
We consider foreground-reduced Q (41 GHz), V (61 GHz), and W (94 GHz) band maps from WMAP 9yr data release that have same \healpix\ pixel resolution of \nside=512 but different beam smoothing levels. 
Then, we also use \planck\ PR4 frequency-specific CMB temperature maps at 70, 100, 143, and 217~GHz channels processed through the \sevem\ component separation method~\cite{sevem2008,sevem2012}. The \sevem\ 70~GHz map is provided at \nside=1024 while the later three (i.e., 100, 143 and 217~GHz) maps are made available at \nside=2048. All of these also have different beam resolutions.

Owing to different resolutions of WMAP and \planck\ detectors that observed the microwave sky at different frequencies, we bring all these different data maps to a common beam and pixel resolution. We do so by processing each map in harmonic space as
\begin{equation}
 a_{l m}^{\rm out} = \left(\frac{{b_l^{\rm out}} p_l^{\rm out}}{{b_l^{\rm in}}p_l^{\rm in}} \right) a_{l m}^{\rm in} 
 \label{eq:downgrade-map}
\end{equation}
where $a_{l m}$ are the spherical harmonic coefficients, `$b_l$' represents the circularized beam transfer function of a particular instrument, and `$p_l$' denotes pixel window function corresponding to the \nside\ parameter of a map digitized using \healpix. The superscript `in' and `out' stands for the input and output map properties, respectively.

We use a Gaussian beam transfer function with FWHM=$30'$ to smooth all frequency-specific maps. Further all are synthesized at \nside=512. Therefore, $b_l^{\rm out}=b_l^{30'}$ and $p_l^{\rm out}=p_l^{512}$.
WMAP team has provided $b_l$'s of each differencing assembly of a frequency channel separately. We used average beam transfer functions corresponding to Q, V and W band maps\footnote{There are 2, 2 and 4 differencing assemblies (DAs) corresponding to Q, V and W bands of WMAP. However in case of Q-band, the mean $b_l$ showed an unphysical upward trend beyond $l = 930$. So, we used average Q-band $b_l$ up to $l = 930$ as given by WMAP team, and extrapolated it beyond that using a quadratic spline interpolator employing \Scipy's \texttt{interp1d} function.}. Since WMAP provided these maps at \nside=512, there is no need for pixel window correction (as $p_l^{\rm in} =p_l^{\rm out} = p_l^{512}$)\footnote{WMAP data maps and beam window functions are available from NASA's Legacy Archive for Microwave Background Data Analysis (LAMBDA) site : \url{https://lambda.gsfc.nasa.gov/product/wmap/current/index.html}.}.
The \sevem\ cleaned frequency-specific CMB maps at 70, 100, 143 and 217~GHz channels also have different beam resolutions as a result of different detector properties. These maps are available from \planck\ legacy archive\footnote{\url{http://pla.esac.esa.int/pla}} (PLA). The individual beam transfer functions are available from \texttt{NERSC}\footnote{CMB data/simulations from various missions, current and planned, are available at \texttt{NERSC} supercomputing facility. More details on \planck\ PR4 (that used \texttt{NPIPE} data processing pipeline~\cite{plk2020npipe}) can be found at the web link : \url{https://wiki.cosmos.esa.int/planck-legacy-archive/index.php/NPIPE_Introduction}. The beam files of different frequency channels are available in the folder \texttt{/global/cfs/cdirs/cmb/data/planck2020} on \texttt{NERSC}.}.

All seven frequency-specific cleaned CMB maps used in the present work are shown in Fig.~[\ref{fig:datamaps}]. The \emph{top row} depicts WMAP's 9yr Q and V band foreground-reduced maps. Depicted in \emph{second row} are WMAP's 9yr foreground-reduced W-band map and \sevem\ cleaned 70~GHz map from \planck\ PR4. Then, \sevem\ 100 and 143~GHz cleaned \planck\ PR4 maps are shown in the \emph{third row}. Finally, shown at the \emph{bottom} is the \sevem\ cleaned 217~GHz map from \planck\ public release 4. All maps are at \nside=512 with a beam smoothing level of a Gaussian window of FWHM=$30'$.

\subsection{Simulations}
\label{sec:sim}
To assess the statistical significance of hemispherical power asymmetry seen in observations, we made use of simulated CMB realizations corresponding to both WMAP and \planck\ datasets.

In case of WMAP, we generated 1000 simulated CMB temperature anisotropy maps for each frequency band viz., Q, V, and W bands from theoretical power spectrum using the \verb+synfast+ module of \healpix, that is then convolved with corresponding (average) beam transfer functions (i.e., $b_l^Q$, $b_l^V$ and $b_l^W$ as described in Sec.~\ref{sec:data}). This theoretical power spectrum is derived based on WMAP's 9yr best-fit cosmological parameters of the $\Lambda$CDM model using \texttt{CAMB}\footnote{\url{https://github.com/cmbant/CAMB}}. 
WMAP satellite scanned the CMB sky multiple times, and correspondingly an $N_{\rm obs}(p)$ map is provided that informs us of the number of times a particular pixel `$p$' (i.e., sky location) is visited during its complete observing period. 
Therefore, we generate noise realizations using the per-pixel observation count `$N^i_{\rm obs}(p)$' and the noise RMS `$\sigma^i_0$' provided by WMAP team for each frequency band `$i$' as
\begin{equation}
n_{\rm noise}^{i} (p) = \frac{\sigma_0^i}{\sqrt{N_{\rm obs}^i(p)}} \times \mathrm{GRN}(0,1)\,.
\label{eq:noise-WMAP}
\end{equation}
Here $\mathrm{GRN}(0,1)$ denotes a Gaussian random number with mean zero and unit variance. The values of $\sigma_0^i$ as given by WMAP team for $i=$Q, V, and W bands are 2.188, 3.131, and 6.544 (in milli-Kelvin), respectively.

Each of the 1000 simulated CMB maps convolved with frequency-specific beam $b_l^i$ (where $i$=Q,V,W) is then combined with an appropriate noise realization (generated following Eq.~\ref{eq:noise-WMAP}). These are then processed similar to data to derive the final set of simulated Q, V and W band maps at \nside=512 with a Gaussian beam whose FWHM=$30'$.

For the analysis of \planck\ PR4 maps, we made use of 600 frequency-specific complementary CMB realizations processed using the \sevem\ pipeline and have the same beam and pixel resolution properties as data. Along with the signal maps, frequency-specific instrumental noise simulations were also provided at the \texttt{NERSC} repository.
These noise realizations are added to the isotropic frequency-specific \sevem\ simulations to replicate the noise properties present in the actual \planck\ data. These noisy CB realizations are downgraded to \nside=512 to have a beam resolution of FWHM=$30'$ Gaussian using the same harmonic space scheme described in Sec.~\ref{sec:data} following Eq.~(\ref{eq:downgrade-map}).
(Here we refer to \planck's FFP12/PR4 simulations as ``isotropic'' for brevity to mean that they are ``unmodulated''. But they contain lensing, Doppler boosting, noise and other observational/instrumentation artifacts that are anisotropic in nature.)

These simulation sets serve as our reference ensembles:
\begin{itemize}
    \item For WMAP data, 1000 simulations corresponding to each of the WMAP's 9yr foreground-reduced Q (41~GHz), V (61~GHz) and W (94~GHz) band maps are employed to determine the statistical significance of HPA detected in the Q, V, and W band data.
    \item For \planck\ derived CMB maps, 600 \sevem\ FFP12 realizations corresponding to each of the \planck\ PR4 \sevem\ cleaned 70, 100, 143 and 217~GHz frequency maps are used to assess HPA in the \planck\ PR4 frequency-specific data.
\end{itemize}

Thus a total of $3\times1000+4\times600=5400$ simulated noisy CMB maps were used in the present work complementing the \emph{seven} frequency-specific cleaned CMB maps.

\subsection{Galactic mask}

WMAP and \planck\ collaborations used different methods to clean raw satellite data for extracting the CMB signal. However, these cleaned CMB maps, when inspected carefully, still possess residual contamination which can bias our results in any cosmological analysis. So, we should mask the contaminated regions in the recovered CMB sky before using it. For our analysis, we employed the WMAP 9yr KQ85 mask ~\cite{Bennett_2013} and the \planck\ PR3 common mask ~\cite{plk18CompSep}. The former is available at  NASA's LAMBDA website and the latter is available at \planck\ legacy archive (PLA). These are binary masks with pixel values either $0$ or $1$ that omit regions from the data where the recovered CMB signal is unreliable.

\begin{figure}[t]
\centering
\includegraphics[width=0.45\textwidth]{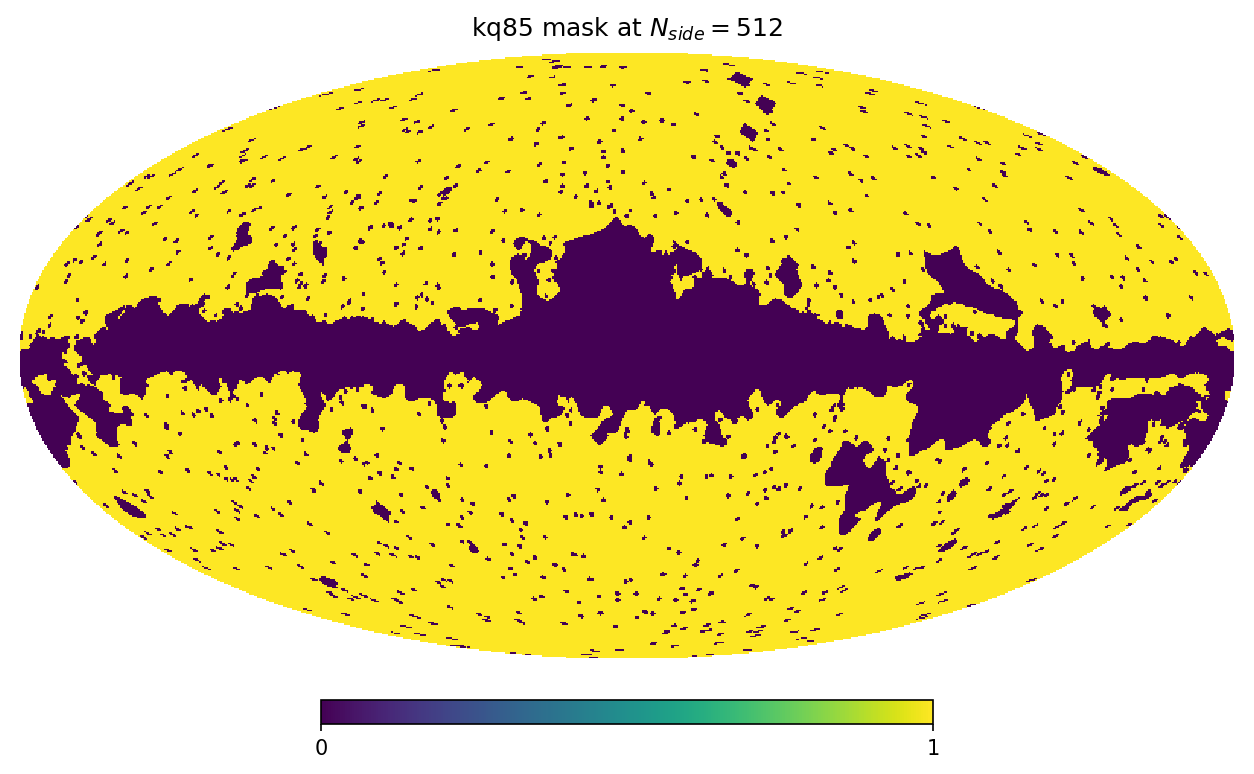}
~
\includegraphics[width=0.45\textwidth]{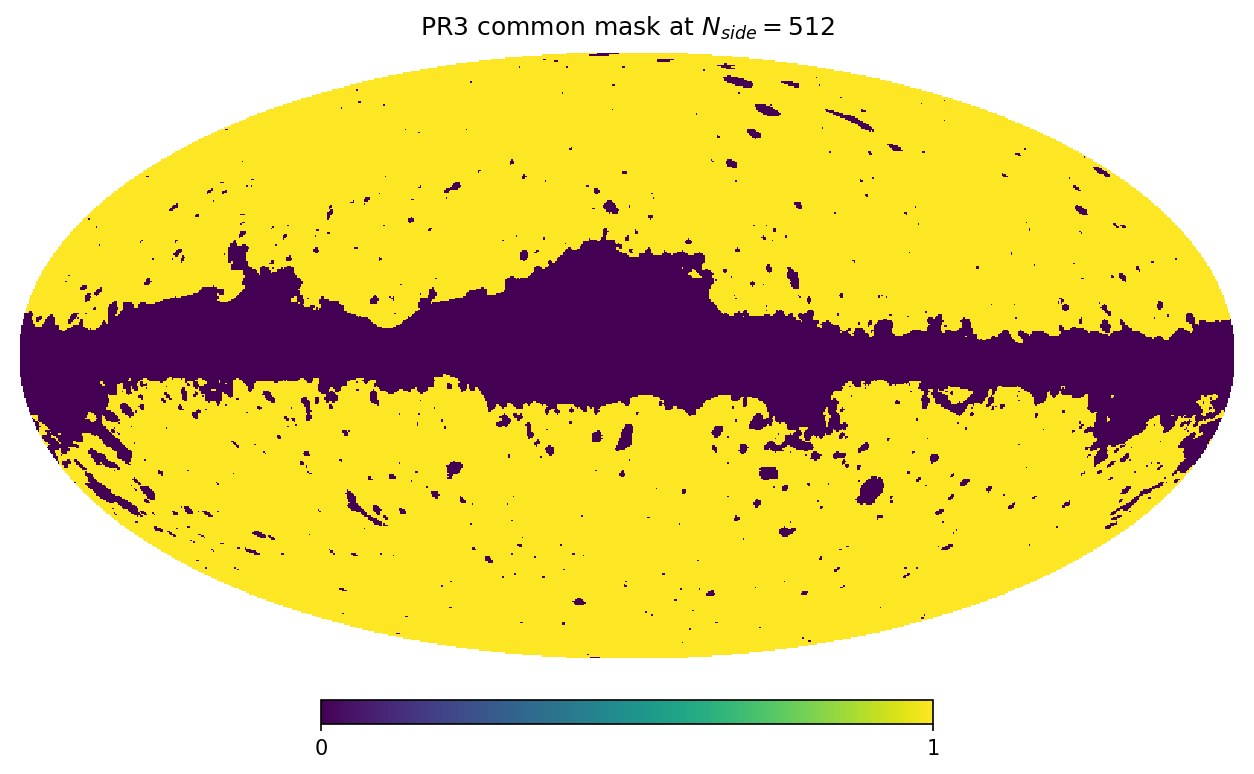}
\includegraphics[width=0.6\textwidth]{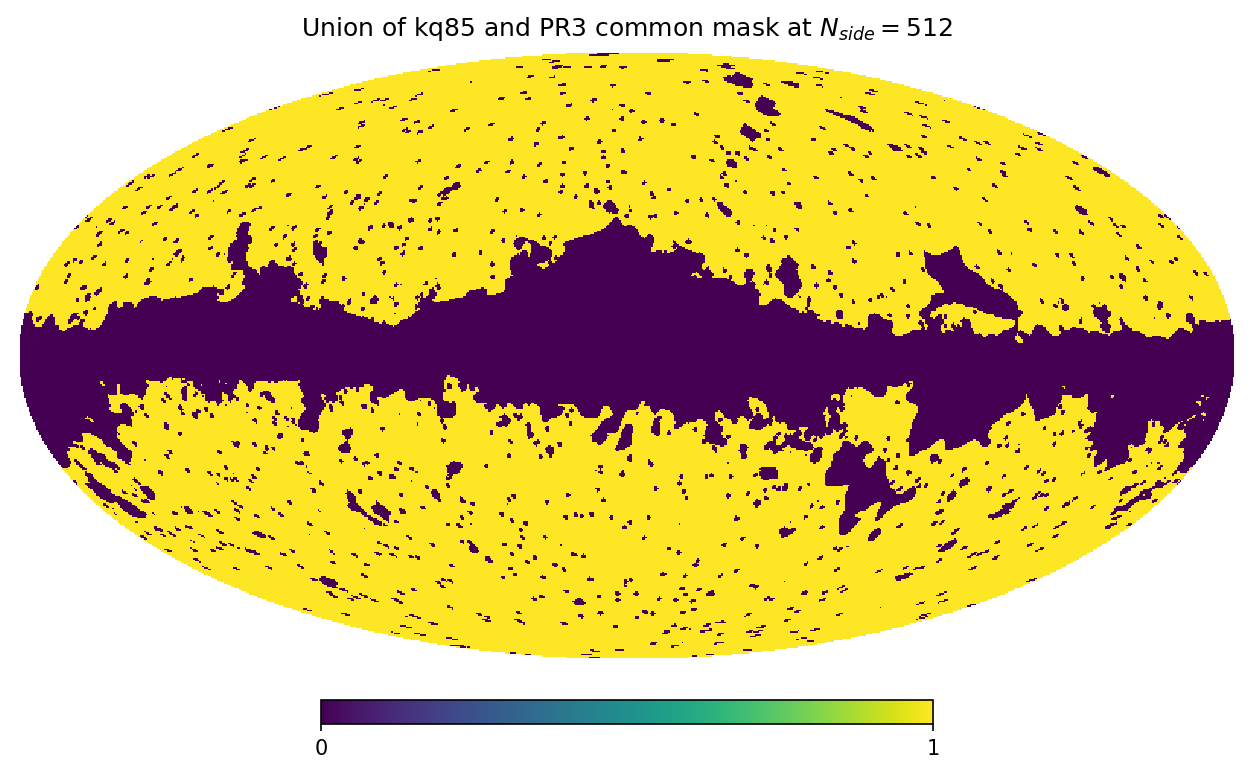}
\caption{The KQ85 mask from WMAP 9yr data release (\emph{top left}), the \planck\ PR3 common mask downgraded to \nside = 512 (\emph{top right}), and their union (\emph{bottom}) which is used commonly on all foreground-reduced CMB maps considered in this work are shown here.}
\label{fig:mask}
\end{figure}

KQ85 mask from WMAP 9yr data release is provided at \nside = 512 with a non-zero sky fraction of  $f_{\mathrm{sky}} \approx 74.80\%$, whereas the \planck\ PR3 common mask is available at \nside=2048 with $f_{\mathrm{sky}} \approx 77.25\%$. So we need to downgrade the PR3 common mask to \nside=512. This is achieved by first inverting the PR3 mask and then applying the same downgrading procedure as used for the PR4 CMB temperature maps outlined following Eq.~(\ref{eq:downgrade-map}). The downgraded mask's $a_{lm}$s thus obtained are synthesized into a map and then thresholded such that those pixels $<0.25$ are set to `0', while the rest are set to `1'. We do this to account for the smearing of point / extended sources due to smoothing. This downgraded, thresholded mask is then inverted to obtain PR3 galactic mask at \nside=512.

These two masks viz., WMAP's 9yr KQ85 mask and the downgraded \planck\ PR3 mask, both at \nside=512 now, are then combined (multiplied) to derive a final mask for use with any of the \emph{seven} foreground cleaned frequency-specific CMB maps that we analyze. The KQ85, downgraded PR3 mask, and the combination of the two are shown in Fig.~[\ref{fig:mask}] all having \nside=512. The combined mask yields an effective usable sky fraction of $f_{\mathrm{sky}} \approx 70.6\%$.

\section{Results from analysis of frequency-specific cleaned CMB maps}
\label{sec:results}

In this section, we present the results from our investigation of the hemispherical power asymmetry (HPA) using WMAP's 9yr and \planck's PR4 frequency-specific CMB temperature maps. Specifically, as stated earlier, we analyze foreground-reduced 41~GHz (Q-band), 61~GHz (V-band) and 94~GHz (W-band) maps from WMAP 9yr data release and \sevem\ cleaned 70~GHz, 100~GHz, 143~GHz and 217~GHz frequency maps from \planck's PR4 in this work to test for any frequency dependence of the HPA anomaly. The analysis results that follow provide insights into the frequency (in)dependence, scale (in)dependence, and robustness of the HPA signal.

We use the local variance estimator (LVE) method described in Sec.~\ref{sec:methods} to probe HPA by deriving local variance maps (LVE maps/LVMs) from observed CMB sky at different frequencies.
LVMs are derived for different choices of disc radii ranging from 1$^{\circ}$ to 40$^{\circ}$ following Eq.~(\ref{eq:var}). LVMs are then normalized using mean field bias map (mean LVE map) estimated from (isotropic) simulations following Eq.~(\ref{eq:norm-var-map}).
In our previous work~\cite{sanyal2024hpa}, we studied the procedural aspects of LVE methodology extensively. One aspect that we identified is to use varying \nside\ grid for different choices of disc radius in generating LVE maps, rather than a fixed \nside=16 as originally proposed in Ref.~\cite{Akrami2014}. This minimizes correlations among pixels of an LVE map and allows a better estimate of the underlying dipole amplitude and direction by using the diagonal elements of the covariance matrix  (via inverse variance weighting).
We calculated the covariance matrices from the local variance maps corresponding to each disc radius chosen using 600 \planck\ simulations for each of the \emph{four} frequency-specific \sevem\ cleaned PR4 CMB maps. The same is followed for WMAP 9yr data derived CMB maps where 1000 simulations were used in the estimation of covariance matrices corresponding to LVMs from the \emph{three} foreground-reduced Q, V and W band maps. These matrices show dominant diagonal elements and significantly smaller off-diagonal terms, suggesting minimal inter-pixel correlation in LVMs. Therefore, the inverse of the diagonal elements are used as weights while fitting the (monopole and) dipole to a normalized variance map using \healpix's \verb+remove_dipole+ subroutine. Finally, as mentioned earlier, we demand availability of at least 100 pixels or 50\% of the full circular disc pixels, whichever is greater, to compute variances locally. Otherwise such pixels of LVM grid at whose centers circular discs are defined are set to \healpix\ bad value of `$1.6375\times10^{30}$' and are omitted from extracting dipole information in LVE maps.

\begin{figure}
\centering
\includegraphics[width=0.92\textwidth]{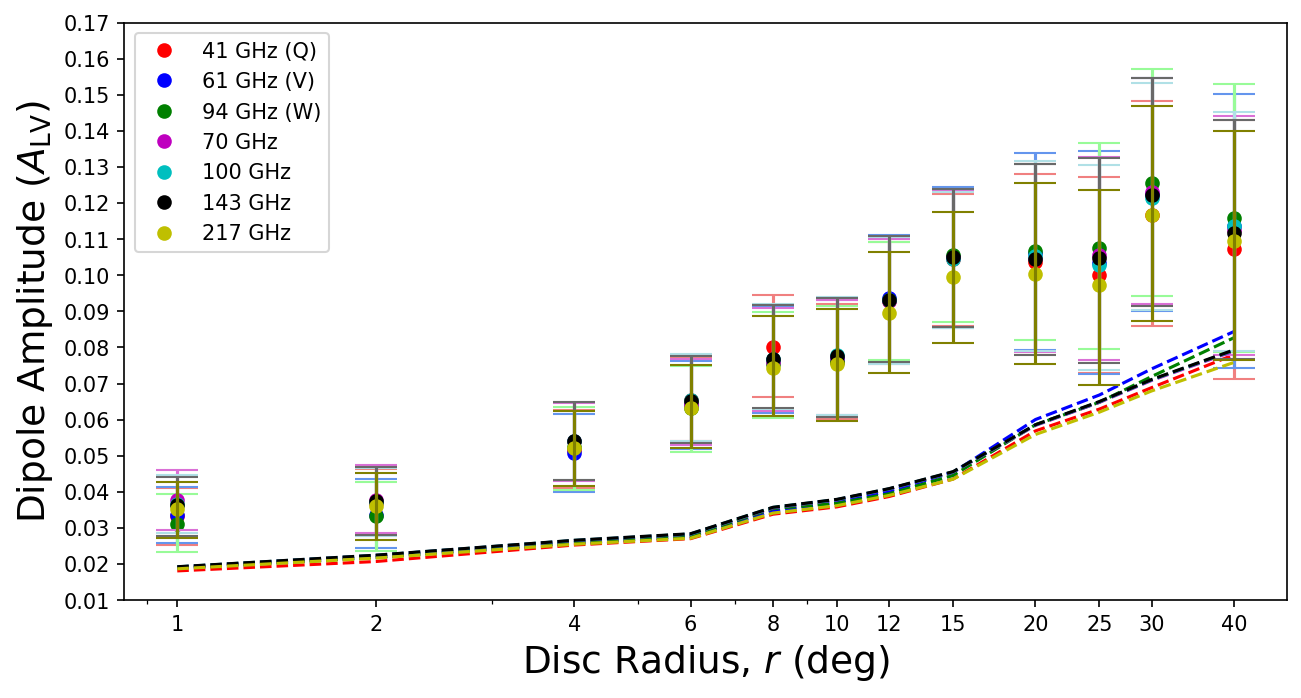}
~
~
\includegraphics[width=0.98\textwidth]{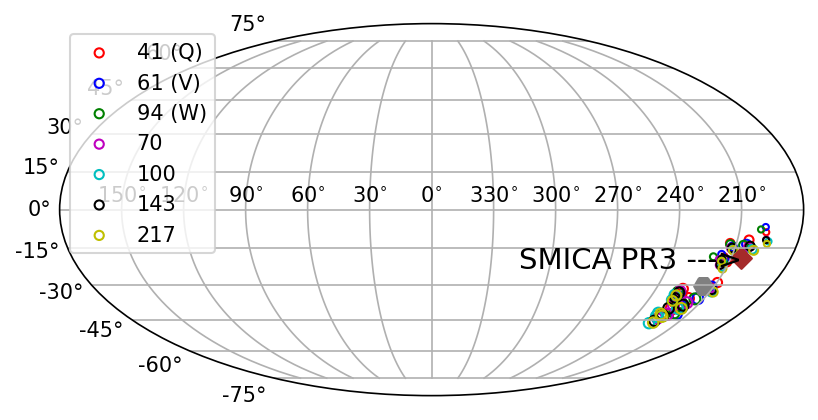}
\caption{\emph{Top :} Dipole amplitudes estimated from the LVE map of different frequency-specific CMB temperature data maps are shown in different colours: \emph{red}, \emph{blue}, \emph{green} are used for
WMAP 9yr Q, V, and W band foreground-reduced CMB maps, whereas \emph{magenta}, \emph{cyan}, \emph{black}, and \emph{yellow} are used
for \planck\ PR4 \sevem\ cleaned 70, 100, 143, and 217~GHz frequency maps, respectively, as a function of disc size `$r$' used in generating LVMs. The error bars are estimated from respective isotropic simulation ensembles. The dashed lines denote expected random dipole amplitude levels in them, that are also estimated from the same set of simulations.
\emph{Bottom :} The dipole directions recovered from WMAP 9yr Q, V, and W band maps, and the \planck\ PR4 \sevem\ cleaned 70, 100, 143, and 217~GHz frequency-specific CMB maps are represented using the same colour scheme as used in the top panel. The LVM dipole directions from different disc radii ($r=1^{\circ}$ to $40^{\circ}$) are shown in open circle point type with increasing point size. The diamond shape in brown colour is the variance asymmetry direction reported by \planck\ team using PR3 {\tt SMICA} cleaned CMB temperature map. The hexagonal shape in gray colour is the average $\approx(215^\circ, -30^\circ)$ of all dipole directions from various frequency-specific temperature maps for all choices of disc sizes.}
\label{fig:dip-ampl-dir-data}
\end{figure}

In the \emph{top panel} of Fig.~[\ref{fig:dip-ampl-dir-data}], we display the dipole amplitudes with error bars corresponding to each of the \emph{seven} frequency-specific foreground-reduced CMB maps considered for this work. Also shown as dashed lines are the expected random dipole amplitude levels in normalized LVMs at different `$r$' for each of these frequency-specific CMB maps from isotropic noisy CMB realizations.
The $1\sigma$ error bars shown with the data are also obtained from the same isotropic simulations corresponding to each CMB map. Further, these error bars are the 16th and 84th quantile values from simulations as it is know that the square of the amplitude (power) follows an asymmetric distribution (a $\chi^2_k$ distribution with $k=3$ degrees of freedom).  
Data points (with error bars) corresponding to WMAP Q (41~GHz), V (61~GHz), and W (94~GHz) bands are depicted in \emph{red}, \emph{blue}, and \emph{green} color, respectively. And the same corresponding to \planck\ PR4 \sevem\ 70, 100, 143, and 217~GHz CMB maps are depicted in \emph{magenta}, \emph{cyan}, \emph{black}, and \emph{yellow}, colours respectively. The same colouring scheme is used to denote the expected random dipole amplitude levels. The observed variation in dipole amplitude with disc radius `$r$' suggests that the assumption of a scale-independent dipole amplitude does not hold. We will test this later in the paper.

\begin{figure}
\centering
\includegraphics[width=0.98\textwidth]{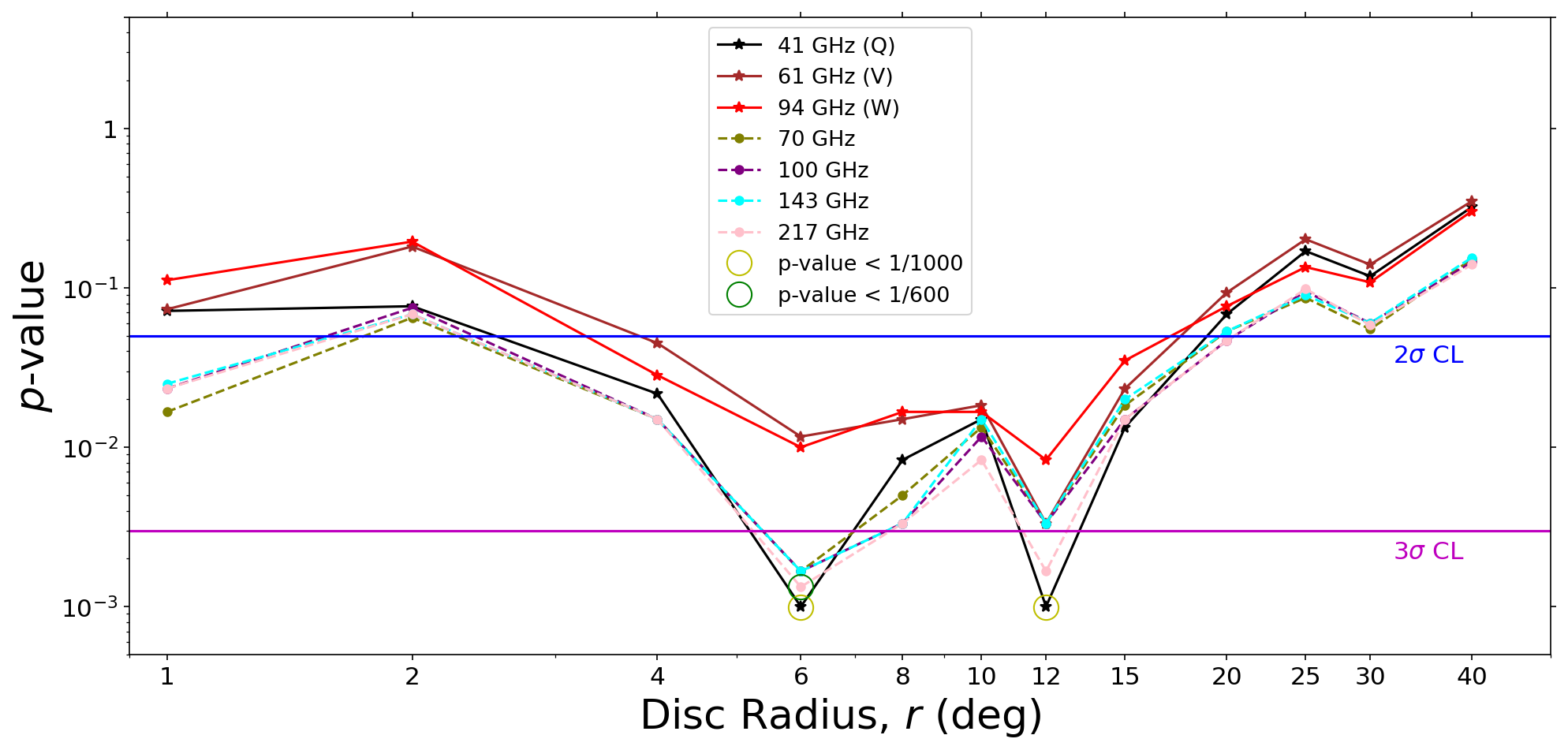}
\caption{$p$-values obtained from various frequency-specific cleaned CMB maps as indicated in the legend are plotted against disc radii. The \emph{yellow} and \emph{green} open circle highlight those disc sizes for which $p$-values are $<1/N_{\rm sim}$ as no simulations are found to have an LVM dipole amplitude that exceeds the data value. Here, the number of simulations $N_{\rm sim}=$1000 and 600, respectively, complementing the WMAP's (Q, V, and W band) and \planck's \sevem\ cleaned PR4 (70, 100, 143, and 217~GHz channel) maps.}
\label{fig:p-vals-all}
\end{figure}

The \emph{bottom panel} of the same figure, Fig.~[\ref{fig:dip-ampl-dir-data}], illustrates the dipole directions recovered from LVE maps of each frequency-specific CMB map. Same colour scheme, as in case of amplitudes shown in the top panel of this figure, is used to denote the directions recovered from different foreground-reduced CMB maps. Open circles with varying point size are representative of the different disc radii employed in our analysis. Large/small disc radii are denoted by large/small open circle point types, respectively, in different colours corresponding to various maps studied as mentioned above.
The recovered LVM dipole directions from different `$r$' are broadly consistent with each other across CMB maps from different frequencies. They cluster in the same region as reported in the literature previously, with little variation. For reference, a \emph{brown} diamond point is shown in the same plot that corresponds to HPA direction found in {\tt SMICA} cleaned CMB map from \planck\ public release 3 (PR3)~\cite{plk2018isostat}. These directions are distributed similar to those reported in our earlier study (using composite \sevem\ cleaned CMB map from \planck\ PR4)~\cite{sanyal2024hpa}.
However, whatever variation that we see in the recovered HPA directions when employing different disc sizes has an interesting pattern. LVM dipole directions from various CMB maps corresponding to larger disc sizes are located away from the galactic plane and those corresponding to smaller disc sizes move towards the galactic plane. The same pattern was found in our previous work also~\cite{sanyal2024hpa}. We remind that various disc sizes probe different angular scales of the CMB sky. The outcome of this LVE analysis is that the observed HPA is not an artifact of instrumental systematics but a persistent and consistent feature across multiple frequency channels and instruments.

To estimate the significance of the observed HPA signal in various maps analyzed, we compute $p$-values using ensembles of isotropic simulations viz., 1000 each for the WMAP 9yr foreground-reduced frequency maps, and 600 simulations each for all \planck\ PR4 channel-specific CMB maps. The combined  results are presented in Fig.~[\ref{fig:p-vals-all}] where $p$-values as a function of disc radius are shown. These $p$-values are computed in a frequentist manner. Solid lines connecting different points in \emph{black}, \emph{brown}, and \emph{red} colour depict the $p$-values across disc sizes for WMAP's 9yr Q, V, and W band maps respectively, whereas dashed lines joining different points in \emph{olive}, \emph{violet}, \emph{cyan}, and \emph{purple} colour are used to depict the $p$-value profile for LVM dipole amplitudes from \planck\ PR4 \sevem\ 70, 100, 143, and 217~GHz CMB maps respectively. The two horizontal lines in \emph{blue} and \emph{magenta} represent the $2\sigma$ and $3\sigma$ confidence levels (CL) respectively.
For a range of disc radii, $r=4^\circ$ to $15^\circ$, all ``seven'' CMB maps studied reveal that HPA is anomalous at $2\sigma$ CL or better, even crossing the $3\sigma$ CL for some disc radii. Also for some disc sizes chosen, the $p$-values approach zero, meaning that none of the simulations exhibit dipole amplitudes larger than that observed in the data. They are highlighted by an \emph{yellow} and \emph{green} open circle for WMAP and \planck\ maps, respectively, where $p$-values are $<1/N_{\rm sim}$ ($N_{\rm sim}$=number of simulations in each ensemble). Overall, the significance with which the HPA feature is found in frequency-specific CMB maps is consistent across all of them.

\begin{figure}[t]
\centering
\includegraphics[width=0.75\textwidth]{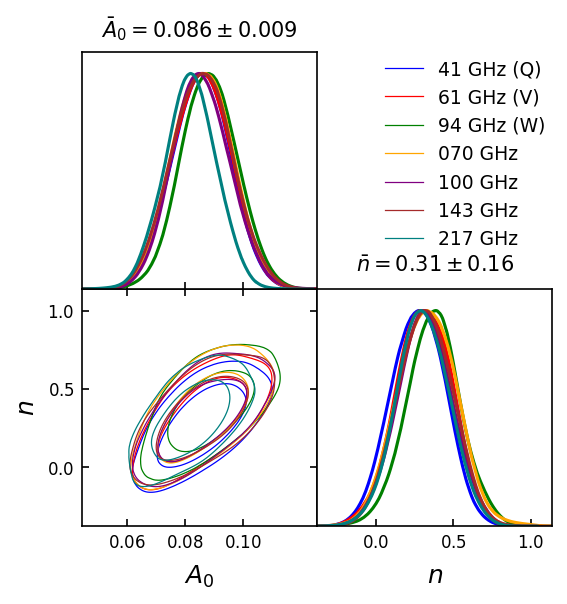}
\caption{Likelihood 2D contour plots along with posteriors of the parameters `$A_{0}$' and `$n$' to fit the scale dependence of HPA (per Eq.~\ref{eq:pl}) are shown with contours signifying $1\sigma$ and $2\sigma$ confidence intervals. The distributions agree well across all cleaned frequency channel maps. See text for more details.} 
\label{fig:hpa-A-n-fit}
\end{figure}

Earlier we noted that the dipole amplitudes recovered from different CMB maps all showed a dependence with disc sizes. Now, to quantify this \emph{angular scale} or equivalently \emph{multipole} dependence of the dipole amplitudes, we performed a power-law fitting to the \emph{corrected} dipole amplitudes from LVE maps following Eq.~(\ref{eq:alv-bias-corr}). As mentioned earlier, the expected bias correction levels for different CMB maps, shown as dashed lines in Fig.~[\ref{fig:dip-ampl-dir-data}], are estimated from isotropic noisy CMB simulations. Further, as identified in Ref.~\cite{sanyal2024hpa}, we consider the range $r\approx 4^\circ$ to $40^\circ$ to do this fitting which is the reliability range for the LVE methodology for the oft quoted amplitude of $A\approx 0.07$ as demonstrated in Ref.~\cite{sanyal2024hpa}.

The disc sizes `$r$' are converted to multipoles using the relation $\ell\sim180^\circ/r$. Following Eq.~(\ref{eq:pl}), a power-law is fit to $A_{\rm LV,corr} \equiv A(\ell)$ for the range $r=6^\circ$ to $30^\circ$ that maps to $\ell=6$ to 30. Treating `$A_{0}$' and `$n$' as free parameters, their best-fit values are determined by performing a Bayesian inference through MCMC sampling (see Appendix~\ref{apdx:pl-fit} for more details). The 2D corner (contour) plots of these parameters obtained from the MCMC fit depicting $1\sigma, 2\sigma$ levels are shown in Fig.~[\ref{fig:hpa-A-n-fit}] along with their posterior distributions. These parameter pairs are shown for all the \emph{seven} foreground cleaned frequency-specific CMB maps from the final WMAP and \planck\ data releases. It is obvious from the plot that the posteriors from different frequency channels are consistent with each other with an average best fit value of $\bar{A}_0 = 0.086 \pm 0.009$ for the amplitude at the pivot scale $\ell_0=10$, and average spectral index $\bar{n} = 0.31 \pm 0.16$. The non-zero value of spectral index `$n$' suggests that the hemispherical power asymmetry is not same over different angular scales, but decreases in strength at small disc radius `$r$' or equivalently higher multipoles `$\ell$'.
Specific values of the parameters $A_0$ and $n$ are presented in Table~\ref{tab:powerlaw_params} for the ``seven'' frequency-specific cleaned CMB maps that we studied in the present work. It is evident from the table that the two parameters are consistent across frequencies.

We note, however, that these amplitudes `$A_0$' are higher than the same value obtained by us earlier using composite \sevem\ cleaned CMB map from \planck\ PR4~\cite{sanyal2024hpa} (that was also smaller than the same values quoted by others who considered scale dependence of HPA dipole amplitude as a power-law~\cite{Tarun2019anom,scal_dep_dm_l23_2019}). As reasoned in our earlier work, this could be because of isotropic CMB modes at high-$l$, among others, that might be diluting the HPA signal being estimated. Here we might be seeing a partial justification to that reasoning in the sense that, by evaluating HPA using relatively low resolution CMB maps viz., at \nside=512 (with $l_{\rm max}=2\times N_{\rm side}=1024$) instead of \nside=2048 (with mode information up to $l=4000$) we are omitting higher multipoles that are isotropic (=unmodulated) and also have noise (specifically in WMAP maps where noise dominates beyond $l\sim800$). So this is resulting in a higher value for `$A_0$' while `$n$' remains essentially same. We however note an interesting pattern for the spectral index `$n$'. It is relatively steeper (large) at $\sim94$~GHz (which is the regime of foreground minimum) compared to other frequencies where `$n$' is relatively flatter (small) where synchrotron/free-free dominate at low frequencies and thermal dust dominates at higher frequencies.

Individual power-law fits to (bias corrected) dipole amplitudes as a function of angular scale ($\ell$) from various frequency-specific foreground cleaned CMB maps listed in Table~\ref{tab:powerlaw_params} are presented in Fig.~[\ref{fig:apdx:power_law_par}] in Appendix~\ref{apdx:pl-fit}.

\begingroup

\renewcommand{\arraystretch}{1.25} 
\begin{table}
\centering
\begin{tabular}{ccl}
\hline
CMB map & $A_0$ & \qquad $n$ \\
\hline
WMAP 9yr Q-band (41~GHz)    & $0.086 \pm 0.009$ & $0.27 \pm 0.16$ \\
WMAP 9yr V-band (61~GHz)    & $0.085 \pm 0.009$ & $0.31 \pm 0.16$ \\
\planck\ PR4 \sevem\ 70~GHz & $0.086 \pm 0.010$ & $0.33 \pm 0.18$ \\
WMAP 9yr W-band (94~GHz) & $0.088 \pm 0.009$ & $0.35_{-0.16}^{+0.18}$ \\
\planck\ PR4 \sevem\ 100~GHz & $0.086 \pm 0.009$ & $0.32 \pm 0.16$ \\
\planck\ PR4 \sevem\ 143~GHz & $0.086 \pm 0.010$ & $0.31 \pm 0.16$ \\
\planck\ PR4 \sevem\ 217~GHz & $0.082 \pm 0.008$ & $0.30 \pm 0.16$ \\
\hline
\end{tabular}
\caption{Posterior mean and $1\sigma$ uncertainties for the power-law fit parameters `$A_{0}$' and `$n$' for the form of the scale dependence considered in the present work (per Eq.~\ref{eq:pl}). For these fits, bias corrected dipole amplitudes from LVE maps, following Eq.~(\ref{eq:alv-bias-corr}), are used from disc sizes $r = 6^\circ$ to $30^\circ$, or analogously the multipoles $\ell=6$ to 30 (following the relation $\ell\sim180^\circ/r$).}
\label{tab:powerlaw_params}
\end{table}

\endgroup

\section{Conclusions}
\label{sec:concl}

In this paper, we presented our investigation on one of the prominent CMB anomalies viz., the hemispherical power asymmetry (HPA) using a pixel-space estimator, namely the local variance estimator (LVE), as applied to several frequency-specific CMB temperature maps.
We explored the frequency dependence of the HPA using these individual frequency-band cleaned CMB maps from WMAP 9yr data release and \planck\ public release 4. We chose the foreground-reduced Q (41~GHz), V (61~GHz), W (94~GHz) band maps from  WMAP 9yr data release and the \planck\ \sevem\ cleaned 70~GHz, 100~GHz, 143~GHz, and 217~GHz frequency maps, making them a total \emph{seven} CMB maps derived at wide ranging frequency channels.
Circular discs of various sizes are defined locally that uniformly covered the entire CMB sky to probe HPA by angular scale by computing local variance maps (LVE maps/LVMs) from a given CMB map. A wide range of disc radii $r=1^\circ$ to $40^\circ$ were considered in this study to derive LVMs.
Our focus here is to quantify the consistency of HPA feature (such as a dipole modulation of otherwise isotropic CMB sky) across different frequency-specific cleaned CMB maps to gauge its cosmic nature and to further assess any angular scale (multipole) dependence of HPA's strength (i.e., dipole amplitude).

Since the CMB maps from different frequency bands and missions are provided at different resolutions, all are processed to have same beam smoothing given by a Gaussian window of FWHM=$30'$ (arcmin) and pixel resolution of \healpix\ \nside=512. To complement the data, 1000 simulations were generated to mimic WMAP's 9yr Q, V, and W band foreground-reduced maps, while \planck\ collaboration provided 600 FFP12 simulation ensembles corresponding to each of the \sevem\ cleaned PR4 channel maps. All these simulations are generated/processed starting with their native resolution (as made available) similar to data, to synthesize ``seven'' sets of isotropic noisy CMB realizations matching the final data maps' properties common to all as mentioned above.
The statistical significance of the observed HPA feature in different frequency channels for different choices of disc sizes via LVE maps is assessed by comparing the data derived dipole amplitudes in respective LVMs to those obtained from complementary simulations.

Our LVE analysis reveals that the WMAP 9yr foreground-cleaned Q (41~GHz), V (61~GHz), W (94~GHz) band maps exhibit an anomalous dipolar component in the local variance maps. Despite relatively higher noise levels in WMAP maps compared to \planck\ maps, the dipole amplitudes remain consistently higher than the expected random dipole strength in them (even if the maps are statistically isotropic). The same dipole component is seen in the LVMs computed from the \planck\ PR4 \sevem-processed frequency-specific CMB temperature data. For all four frequency channels (70, 100, 143, and 217~GHz), the dipole amplitude of the variance map exceeds the expected values at different `$r$' found in simulations. For disc sizes of $r=4^\circ$ to $\lesssim20^\circ$ the HPA in all the ``seven'' CMB maps studied, lies outside the $2\sigma$ level, even crossing the $3\sigma$ level for some disc radii. In some cases, none of the simulations derived LVE maps' dipole amplitudes exceed the observations. The direction of the dipole is also consistent in all LVMs from across frequency maps and lies close to the previously reported HPA direction in the literature. There is however, some variation to note with the disc radius employed in computing LVMs, though very clustered in all the CMB maps studied.
Thus HPA is consistent in amplitude and direction across CMB maps derived from different frequencies fortifying its robustness between experiments and instruments.

In order to assess the scale dependence of HPA, we fit a power-law model (per Eq.~\ref{eq:pl}) between observed dipole amplitudes (that are corrected for expected random dipole component per Eq.~\ref{eq:alv-bias-corr}) and multipole `$\ell$' obtained following the relation $\ell \sim 180^\circ/r$. The best-fit amplitudes are found to be $A_0 \approx 0.086$ from all the ``seven'' frequency-specific cleaned CMB maps at the pivot scale of $\ell_0=10$, and a non-zero value of the index $n \approx 0.3$ (with specific values given in Table~\ref{tab:powerlaw_params}). This hints at a consistent striking deviation from a scale independent HPA phenomena across angular scales, contrary to what it is usually assumed to be. (Even though HPA is understood to be a low multipole anomaly up to $l\lesssim64$, its amplitude is assumed to be constant $A\sim0.07$ across all these modes, akin to a step function that vanishes at $l\gtrsim64$.) 

So, the hemispherical power asymmetry seen in CMB sky is independent of frequency, as it is found to be present across all \emph{seven} frequency-specific cleaned CMB maps from different missions with similar strength and direction. This negates the possibility of HPA being of non-cosmic origin, but is a robust feature intrinsic to the CMB (temperature) sky. Further, this hemispherical power asymmetry was also found to vary in strength by angular scale, that is strong at low multipoles and falling off at intermediate to higher modes of CMB sky. Its physical origin is still elusive and remains as an open question to the cosmology community. The study of HPA in polarization data, with future missions that aim to probe CMB polarization signal precisely, will shed more light in solving this puzzle.

\section*{Acknowledgments}
SS acknowledges financial support from the University Grant Commission, India through UGC-Ref No.:1487/(CSIR-UGC NET JUNE 2018). We acknowledge the use of NASA's WMAP data from the Legacy Archive for Microwave Background Data
Analysis (LAMBDA), part of the High Energy Astrophysics Science Archive Center (HEASARC). HEASARC/LAMBDA is a service of the Astrophysics Science Division at the NASA Goddard Space Flight Center.
Part of the results presented here are based on observations obtained with \planck\, an ESA science mission with instruments and contributions directly funded by ESA Member States, NASA, and Canada.
This research used resources of the National Energy Research Scientific Computing Center (NERSC), a U.S. Department of Energy Office of Science User Facility operated under Contract No. DE-AC02-05CH11231.
Further, this work also made use of the following software packages :
\texttt{Cobaya}\footnote{\url{https://cobaya.readthedocs.io/en/latest/}}~\cite{Cobaya2021}, \texttt{GetDist}\footnote{\url{https://getdist.readthedocs.io/en/latest/}}~\cite{GetDist2019},
{\tt HEALPix}/{\tt Healpy}~\cite{healpix,healpy},
\texttt{SciPy}\footnote{\url{https://scipy.org}}~\cite{scipy2020},
\texttt{NumPy}\footnote{\url{https://numpy.org}}~\cite{numpy2020},
\texttt{Astropy}\footnote{\url{http://www.astropy.org}}~\cite{astropy2013,astropy2018,astropy2022},
and \texttt{matplotlib}\footnote{\url{https://matplotlib.org/stable/index.html}}~\cite{matplotlib}.


\bibliographystyle{unsrt}
\bibliography{ref_hpa_plk20}

@ARTICLE{schwarz2016,
       author = {{Schwarz}, D. J. and {Copi}, C. J. and {Huterer}, D. and {Starkman}, G. D.},
        title = "{CMB anomalies after Planck}",
      journal = {Class. Quantum Grav.},
         year = {2016},
       volume = {33},
          eid = {184001},
        pages = {184001},
          doi = {10.1088/0264-9381/33/18/184001},
archivePrefix = {arXiv},
       eprint = {1510.07929},
 primaryClass = {astro-ph.CO},
       adsurl = {https://ui.adsabs.harvard.edu/abs/2016CQGra..33r4001S}
}

@ARTICLE{bull2016,
    author = "Bull, P. and others",
    title = "{Beyond {\ensuremath{\Lambda}} CDM: Problems, solutions, and the road ahead}",
    eprint = "1512.05356",
    archivePrefix = "arXiv",
    primaryClass = "astro-ph.CO",
    doi = "10.1016/j.dark.2016.02.001",
    journal = "{Phys. Dark Univ.}",
    volume = "{12}",
    pages = "{56-99}",
    year = "{2016}"
}

@ARTICLE{abdalla2022,
    author = "Abdalla, E. and others",
    title = "{Cosmology intertwined: A review of the particle physics, astrophysics, and cosmology associated with the cosmological tensions and anomalies}",
    eprint = "2203.06142",
    archivePrefix = "arXiv",
    primaryClass = "astro-ph.CO",
    reportNumber = "FERMILAB-CONF-22-192-SCD",
    doi = "10.1016/j.jheap.2022.04.002",
    journal = "{J. High Energy Astrophys.}",
    volume = "{34}",
    pages = "{49-211}",
    year = "{2022}"
}

@ARTICLE{cp2023,
       author = "Aluri, Pavan Kumar and others",
        title = "{Is the observable Universe consistent with the cosmological principle?}",
      journal = {Class. Quantum Grav.},
         year = {2023},
       volume = {40},
          eid = {094001},
        pages = {094001},
          doi = {10.1088/1361-6382/acbefc},
archivePrefix = {arXiv},
       eprint = {2207.05765},
 primaryClass = {astro-ph.CO},
       adsurl = {https://ui.adsabs.harvard.edu/abs/2023CQGra..40i4001K}
}

@ARTICLE{healpix,
       author = {{G{\'o}rski}, K.~M. and {Hivon}, E. and {Banday}, A.~J. and {Wandelt}, B.~D. and {Hansen}, F.~K. and {Reinecke}, M. and {Bartelmann}, M.},
        title = "{HEALPix: A Framework for High-Resolution Discretization and Fast Analysis of Data Distributed on the Sphere}",
      journal = {\apj},
         year = 2005,
        month = apr,
       volume = {622},
       number = {2},
        pages = {759-771},
          doi = {10.1086/427976},
       adsurl = {https://ui.adsabs.harvard.edu/abs/2005ApJ...622..759G}
}

@article{healpy,
  doi = {10.21105/joss.01298},
  url = {https://doi.org/10.21105/joss.01298},
  year = {2019},
  month = mar,
  publisher = {The Open Journal},
  volume = {4},
  number = {35},
  pages = {1298},
  author = {Andrea Zonca and Leo Singer and Daniel Lenz and Martin Reinecke and Cyrille Rosset and Eric Hivon and Krzysztof Gorski},
  title = {healpy: equal area pixelization and spherical harmonics transforms for data on the sphere in Python},
  journal = {Journal of Open Source Software}
}

@ARTICLE{scipy2020,
  author  = {Virtanen, Pauli and others},
  title   = {{{SciPy} 1.0: Fundamental Algorithms for Scientific
            Computing in Python}},
  collaboration = "{Astropy Collaboration}",
  journal = {Nature Methods},
  year    = {2020},
  volume  = {17},
  pages   = {261--272},
  adsurl  = {https://rdcu.be/b08Wh},
  doi     = {10.1038/s41592-019-0686-2},
}

@Article{numpy2020,
 title         = {Array programming with {NumPy}},
 author        = {Charles R. Harris and others},
 year          = {2020},
 month         = sep,
 journal       = {Nature},
 volume        = {585},
 number        = {7825},
 pages         = {357--362},
 doi           = {10.1038/s41586-020-2649-2},
 publisher     = {Springer Science and Business Media {LLC}},
 url           = {https://doi.org/10.1038/s41586-020-2649-2}
}

@article{astropy2013,
Adsurl = {http://adsabs.harvard.edu/abs/2013A%26A...558A..33A},
Archiveprefix = {arXiv},
Author = {{Robitaille}, T.~P. and others},
collaboration = "{Astropy Collaboration}",
Doi = {10.1051/0004-6361/201322068},
Eid = {A33},
Eprint = {1307.6212},
Journal = {\aap},
Month = oct,
Pages = {A33},
Primaryclass = {astro-ph.IM},
Title = {{Astropy: A community Python package for astronomy}},
Volume = 558,
Year = 2013}

@ARTICLE{astropy2018,
       author = {{Price-Whelan}, A.~M. and others},
        title = "{The Astropy Project: Building an Open-science Project and Status of the v2.0 Core Package}",
collaboration = "{Astropy Collaboration}",
      journal = {\aj},
         year = 2018,
        month = sep,
       volume = {156},
       number = {3},
          eid = {123},
        pages = {123},
          doi = {10.3847/1538-3881/aabc4f},
archivePrefix = {arXiv},
       eprint = {1801.02634},
 primaryClass = {astro-ph.IM},
       adsurl = {https://ui.adsabs.harvard.edu/abs/2018AJ....156..123A}
}

@ARTICLE{astropy2022,
       author = {{Price-Whelan}, Adrian M. and others},
        title = "{The Astropy Project: Sustaining and Growing a Community-oriented Open-source Project and the Latest Major Release (v5.0) of the Core Package}",
collaboration = "{Astropy Collaboration}",
      journal = {\apj},
         year = 2022,
        month = aug,
       volume = {935},
       number = {2},
          eid = {167},
        pages = {167},
          doi = {10.3847/1538-4357/ac7c74},
archivePrefix = {arXiv},
       eprint = {2206.14220},
 primaryClass = {astro-ph.IM},
       adsurl = {https://ui.adsabs.harvard.edu/abs/2022ApJ...935..167A}
}

@Article{matplotlib,
  Author    = {Hunter, J. D.},
  Title     = {Matplotlib: A 2D graphics environment},
  Journal   = {Computing in Science \& Engineering},
  Volume    = {9},
  Number    = {3},
  Pages     = {90--95},
  publisher = {IEEE COMPUTER SOC},
  doi       = {10.1109/MCSE.2007.55},
  year      = 2007
}

@ARTICLE{Cobaya2021,
       author = {{Torrado}, Jes{\'u}s and {Lewis}, Antony},
        title = "{Cobaya: code for Bayesian analysis of hierarchical physical models}",
      journal = {\jcap},
         year = 2021,
        month = may,
       volume = {2021},
       number = {5},
          eid = {057},
        pages = {057},
          doi = {10.1088/1475-7516/2021/05/057},
archivePrefix = {arXiv},
       eprint = {2005.05290},
 primaryClass = {astro-ph.IM},
       adsurl = {https://ui.adsabs.harvard.edu/abs/2021JCAP...05..057T}
}

@ARTICLE{GetDist2019,
       author = {{Lewis}, Antony},
        title = "{GetDist: a Python package for analysing Monte Carlo samples}",
      journal = {arXiv e-prints},
         year = 2019,
        month = oct,
          eid = {arXiv:1910.13970},
        pages = {arXiv:1910.13970},
          doi = {10.48550/arXiv.1910.13970},
archivePrefix = {arXiv},
       eprint = {1910.13970},
 primaryClass = {astro-ph.IM},
       adsurl = {https://ui.adsabs.harvard.edu/abs/2019arXiv191013970L}
}

@ARTICLE{wmap7yranom,
       author = {{Bennett}, C.~L. and others},
        title = "{Seven-year Wilkinson Microwave Anisotropy Probe (WMAP) Observations: Are There Cosmic Microwave Background Anomalies?}",
      journal = {\apjs},
         year = 2011,
        month = feb,
       volume = {192},
       number = {2},
          eid = {17},
        pages = {17},
          doi = {10.1088/0067-0049/192/2/17},
archivePrefix = {arXiv},
       eprint = {1001.4758},
 primaryClass = {astro-ph.CO},
       adsurl = {https://ui.adsabs.harvard.edu/abs/2011ApJS..192...17B}
}

@article{plk2013isostat,
    author = "Ade, P. A. R. and others",
    collaboration = "Planck",
    title = "{Planck 2013 results. XXIII. Isotropy and statistics of the CMB}",
    eprint = "1303.5083",
    archivePrefix = "arXiv",
    primaryClass = "astro-ph.CO",
    reportNumber = "CERN-PH-TH-2013-136",
    doi = "10.1051/0004-6361/201321534",
    journal = {\aap},
    volume = "571",
    pages = "A23",
    year = "2014"
}

@article{plk2015isostat,
    author = "Ade, P. A. R. and others",
    collaboration = "Planck",
    title = "{Planck 2015 results. XVI. Isotropy and statistics of the CMB}",
    eprint = "1506.07135",
    archivePrefix = "arXiv",
    primaryClass = "astro-ph.CO",
    doi = "10.1051/0004-6361/201526681",
    journal = {\aap},
    volume = "594",
    pages = "A16",
    year = "2016"
}

@ARTICLE{plk2018isostat,
    author = "Akrami, Y. and others",
    collaboration = "Planck",
    title = "{Planck 2018 results. VII. Isotropy and Statistics of the CMB}",
    eprint = "1906.02552",
    archivePrefix = "arXiv",
    primaryClass = "astro-ph.CO",
    doi = "10.1051/0004-6361/201935201",
    journal = {\aap},
    volume = "641",
    pages = "A7",
    year = "2020"
}

@ARTICLE{Akrami2014,
       author = {{Akrami}, Y. and {Fantaye}, Y. and {Shafieloo}, A. and {Eriksen}, H.~K. and {Hansen}, F.~K. and {Banday}, A.~J. and {G{\'o}rski}, K.~M.},
        title = "{Power Asymmetry in WMAP and Planck Temperature Sky Maps as Measured by a Local Variance Estimator}",
      journal = {\apjl},
         year = 2014,
        month = apr,
       volume = {784},
       number = {2},
          eid = {L42},
        pages = {L42},
          doi = {10.1088/2041-8205/784/2/L42},
archivePrefix = {arXiv},
       eprint = {1402.0870},
 primaryClass = {astro-ph.CO},
       adsurl = {https://ui.adsabs.harvard.edu/abs/2014ApJ...784L..42A}
}

@ARTICLE{Adhikari2015,
       author = {{Adhikari}, Saroj},
        title = "{Local variance asymmetries in Planck temperature anisotropy maps}",
      journal = {\mnras},
         year = 2015,
        month = feb,
       volume = {446},
       number = {4},
        pages = {4232-4238},
          doi = {10.1093/mnras/stu2408},
archivePrefix = {arXiv},
       eprint = {1408.5396},
 primaryClass = {astro-ph.CO},
       adsurl = {https://ui.adsabs.harvard.edu/abs/2015MNRAS.446.4232A}
}

@article{Gimeno_Amo_2023,
   title={Hemispherical power asymmetry in intensity and polarization for Planck PR4 data},
   volume={2023},
   ISSN={1475-7516},
   url={http://dx.doi.org/10.1088/1475-7516/2023/12/029},
   DOI={10.1088/1475-7516/2023/12/029},
   number={12},
   journal={\jcap},
   publisher={IOP Publishing},
   author={Gimeno-Amo, C. and Barreiro, R.B. and Martínez-González, E. and Marcos-Caballero, A.},
   year={2023},
   month=dec, pages={029}
}

@article{sanyal2024hpa,
       author = {{Sanyal}, Sanjeev and {Patel}, Sanjeet K. and {Aluri}, Pavan K. and {Shafieloo}, Arman},
        title = "{A reassessment of LVE method and hemispherical power asymmetry in CMB temperature data from Planck PR4}",
      journal = {arXiv e-prints},
         year = 2024,
        month = nov,
          eid = {arXiv:2411.15786},
        pages = {arXiv:2411.15786},
          doi = {10.48550/arXiv.2411.15786},
archivePrefix = {arXiv},
       eprint = {2411.15786},
 primaryClass = {astro-ph.CO},
       adsurl = {https://ui.adsabs.harvard.edu/abs/2024arXiv241115786S}
}

@article{Eriksen_2004,
   title={Asymmetries in the Cosmic Microwave Background Anisotropy Field},
   volume={605},
   ISSN={1538-4357},
   url={http://dx.doi.org/10.1086/382267},
   DOI={10.1086/382267},
   number={1},
   journal={\apj},
   publisher={American Astronomical Society},
   author={Eriksen, H. K. and Hansen, F. K. and Banday, A. J. and Gorski, K. M. and Lilje, P. B.},
   year={2004},
   month=apr, pages={14–20}
}

@article{Hansen_2004,
   title={Testing the cosmological principle of isotropy: local power-spectrum estimates of theWMAPdata},
   volume={354},
   ISSN={1365-2966},
   url={http://dx.doi.org/10.1111/j.1365-2966.2004.08229.x},
   DOI={10.1111/j.1365-2966.2004.08229.x},
   number={3},
   journal={\mnras},
   publisher={Oxford University Press (OUP)},
   author={Hansen, F. K. and Banday, A. J. and Górski, K. M.},
   year={2004},
   month=nov, pages={641–665}
}

@article{Gordon2005,
  title = {Spontaneous isotropy breaking: A mechanism for CMB multipole alignments},
  author = {Gordon, Christopher and Hu, Wayne and Huterer, Dragan and Crawford, Tom},
  journal = {\prd},
  volume = {72},
  issue = {10},
  pages = {103002},
  numpages = {13},
  year = {2005},
  month = {Nov},
  publisher = {American Physical Society},
  doi = {10.1103/PhysRevD.72.103002},
  url = {https://link.aps.org/doi/10.1103/PhysRevD.72.103002}
}

@article{Eriksen_2007,
       author = {{Eriksen}, H.~K. and {Banday}, A.~J. and {G{\'o}rski}, K.~M. and {Hansen}, F.~K. and {Lilje}, P.~B.},
        title = "{Hemispherical Power Asymmetry in the Third-Year Wilkinson Microwave Anisotropy Probe Sky Maps}",
      journal = {\apjl},
         year = 2007,
        month = may,
       volume = {660},
       number = {2},
        pages = {L81-L84},
          doi = {10.1086/518091},
archivePrefix = {arXiv},
       eprint = {astro-ph/0701089},
 primaryClass = {astro-ph},
       adsurl = {https://ui.adsabs.harvard.edu/abs/2007ApJ...660L..81E}
}

@ARTICLE{Lew2008,
       author = {{Lew}, Bartosz},
        title = "{Hemispherical power asymmetry: parameter estimation from cosmic microwave background WMAP5 data}",
      journal = {\jcap},
         year = 2008,
        month = sep,
       volume = {2008},
       number = {9},
          eid = {023},
        pages = {023},
          doi = {10.1088/1475-7516/2008/09/023},
archivePrefix = {arXiv},
       eprint = {0808.2867},
 primaryClass = {astro-ph},
       adsurl = {https://ui.adsabs.harvard.edu/abs/2008JCAP...09..023L}
}

@ARTICLE{Bernui2008,
       author = {{Bernui}, Armando},
        title = "{Anomalous CMB north-south asymmetry}",
      journal = {\prd},
         year = 2008,
        month = sep,
       volume = {78},
       number = {6},
          eid = {063531},
        pages = {063531},
          doi = {10.1103/PhysRevD.78.063531},
archivePrefix = {arXiv},
       eprint = {0809.0934},
 primaryClass = {astro-ph},
       adsurl = {https://ui.adsabs.harvard.edu/abs/2008PhRvD..78f3531B}
}

@article{Hoftuft_2009_lowL,
       author = {{Hoftuft}, J. and {Eriksen}, H.~K. and {Banday}, A.~J. and {G{\'o}rski}, K.~M. and {Hansen}, F.~K. and {Lilje}, P.~B.},
        title = "{Increasing Evidence for Hemispherical Power Asymmetry in the Five-Year WMAP Data}",
      journal = {\apj},
         year = 2009,
        month = jul,
       volume = {699},
       number = {2},
        pages = {985-989},
          doi = {10.1088/0004-637X/699/2/985},
archivePrefix = {arXiv},
       eprint = {0903.1229},
 primaryClass = {astro-ph.CO},
       adsurl = {https://ui.adsabs.harvard.edu/abs/2009ApJ...699..985H}
}

@article{Hansen_2009,
       author = {{Hansen}, F.~K. and {Banday}, A.~J. and {G{\'o}rski}, K.~M. and {Eriksen}, H.~K. and {Lilje}, P.~B.},
        title = "{Power Asymmetry in Cosmic Microwave Background Fluctuations from Full Sky to Sub-Degree Scales: Is the Universe Isotropic?}",
      journal = {\apj},
         year = 2009,
        month = oct,
       volume = {704},
       number = {2},
        pages = {1448-1458},
          doi = {10.1088/0004-637X/704/2/1448},
archivePrefix = {arXiv},
       eprint = {0812.3795},
 primaryClass = {astro-ph},
       adsurl = {https://ui.adsabs.harvard.edu/abs/2009ApJ...704.1448H}
}

@ARTICLE{HansonLewis2009,
       author = {{Hanson}, Duncan and {Lewis}, Antony},
        title = "{Estimators for CMB statistical anisotropy}",
      journal = {\prd},
         year = 2009,
        month = sep,
       volume = {80},
       number = {6},
          eid = {063004},
        pages = {063004},
          doi = {10.1103/PhysRevD.80.063004},
archivePrefix = {arXiv},
       eprint = {0908.0963},
 primaryClass = {astro-ph.CO},
       adsurl = {https://ui.adsabs.harvard.edu/abs/2009PhRvD..80f3004H}
}

@article{Paci2010,
    author = {Paci, F. and Gruppuso, A. and Finelli, F. and Cabella, P. and De Rosa, A. and Mandolesi, N. and Natoli, P.},
    title = "{Power asymmetries in the cosmic microwave background temperature and polarization patterns}",
    journal = {\mnras},
    volume = {407},
    number = {1},
    pages = {399-404},
    year = {2010},
    month = {08},
    issn = {0035-8711},
    doi = {10.1111/j.1365-2966.2010.16905.x},
    url = {https://doi.org/10.1111/j.1365-2966.2010.16905.x},
    eprint = {https://academic.oup.com/mnras/article-pdf/407/1/399/3082300/mnras0407-0399.pdf},
}

@ARTICLE{Flender_2013,
       author = {{Flender}, Samuel and {Hotchkiss}, Shaun},
        title = "{The small scale power asymmetry in the cosmic microwave background}",
      journal = {\jcap},
         year = 2013,
        month = sep,
       volume = {2013},
       number = {9},
          eid = {033},
        pages = {033},
          doi = {10.1088/1475-7516/2013/09/033},
archivePrefix = {arXiv},
       eprint = {1307.6069},
 primaryClass = {astro-ph.CO},
       adsurl = {https://ui.adsabs.harvard.edu/abs/2013JCAP...09..033F}
}

@article{Pranati2013,
       author = {{Rath}, Pranati K. and {Jain}, Pankaj},
        title = "{Testing the dipole modulation model in CMBR}",
      journal = {\jcap},
         year = 2013,
        month = dec,
       volume = {2013},
       number = {12},
          eid = {014},
        pages = {014},
          doi = {10.1088/1475-7516/2013/12/014},
archivePrefix = {arXiv},
       eprint = {1308.0924},
 primaryClass = {astro-ph.CO},
       adsurl = {https://ui.adsabs.harvard.edu/abs/2013JCAP...12..014R}
}

@ARTICLE{QuartinNotari2015,
       author = {{Quartin}, Miguel and {Notari}, Alessio},
        title = "{On the significance of power asymmetries in Planck CMB data at all scales}",
      journal = {\jcap},
         year = 2015,
        month = jan,
       volume = {2015},
       number = {1},
        pages = {008-008},
          doi = {10.1088/1475-7516/2015/01/008},
archivePrefix = {arXiv},
       eprint = {1408.5792},
 primaryClass = {astro-ph.CO},
       adsurl = {https://ui.adsabs.harvard.edu/abs/2015JCAP...01..008Q}
}

@article{Aiola_2015,
   title={Microwave background correlations from dipole anisotropy modulation},
   volume={92},
   ISSN={1550-2368},
   url={http://dx.doi.org/10.1103/PhysRevD.92.063008},
   DOI={10.1103/physrevd.92.063008},
   number={6},
   journal={\prd},
   publisher={American Physical Society (APS)},
   author={Aiola, Simone and Wang, Bingjie and Kosowsky, Arthur and Kahniashvili, Tina and Firouzjahi, Hassan},
   year={2015},
   month=sep
}

@article{Ghosh_2016,
   title={Dipole modulation of cosmic microwave background temperature and polarization},
   volume={2016},
   ISSN={1475-7516},
   url={http://dx.doi.org/10.1088/1475-7516/2016/01/046},
   DOI={10.1088/1475-7516/2016/01/046},
   number={01},
   journal={\jcap},
   publisher={IOP Publishing},
   author={Ghosh, Shamik and Kothari, Rahul and Jain, Pankaj and Rath, Pranati K.},
   year={2016},
   month=jan, pages={046–046}
}

@ARTICLE{scal_dep_dm_l23_2019,
       author = {{Marcos-Caballero}, A. and {Mart{\'\i}nez-Gonz{\'a}lez}, E.},
        title = "{Scale-dependent dipolar modulation and the quadrupole-octopole alignment in the CMB temperature}",
      journal = {\jcap},
         year = 2019,
        month = oct,
       volume = {2019},
       number = {10},
          eid = {053},
        pages = {053},
          doi = {10.1088/1475-7516/2019/10/053},
archivePrefix = {arXiv},
       eprint = {1909.06093},
 primaryClass = {astro-ph.CO},
       adsurl = {https://ui.adsabs.harvard.edu/abs/2019JCAP...10..053M}
}

@ARTICLE{Hajian2005,
       author = {{Hajian}, Amir and {Souradeep}, Tarun and {Cornish}, Neil},
        title = "{Statistical Isotropy of the Wilkinson Microwave Anisotropy Probe Data: A Bipolar Power Spectrum Analysis}",
      journal = {\apjl},
         year = 2005,
        month = jan,
       volume = {618},
       number = {2},
        pages = {L63-L66},
          doi = {10.1086/427652},
archivePrefix = {arXiv},
       eprint = {astro-ph/0406354},
 primaryClass = {astro-ph},
       adsurl = {https://ui.adsabs.harvard.edu/abs/2005ApJ...618L..63H}
}

@ARTICLE{Tarun2019anom,
       author = {{Shaikh}, Shabbir and {Mukherjee}, Suvodip and {Das}, Santanu and {Wandelt}, Benjamin D. and {Souradeep}, Tarun},
        title = "{Joint Bayesian analysis of large angular scale CMB temperature anomalies}",
      journal = {\jcap},
         year = 2019,
        month = aug,
       volume = {2019},
       number = {8},
          eid = {007},
        pages = {007},
          doi = {10.1088/1475-7516/2019/08/007},
archivePrefix = {arXiv},
       eprint = {1902.10155},
 primaryClass = {astro-ph.CO},
       adsurl = {https://ui.adsabs.harvard.edu/abs/2019JCAP...08..007S}
}

@ARTICLE{rajib2023hpaai,
       author = {{Khan}, Md Ishaque and {Saha}, Rajib},
        title = "{Detection of Dipole Modulation in CMB Temperature Anisotropy Maps from WMAP and Planck using Artificial Intelligence}",
      journal = {\apj},
         year = 2023,
        month = apr,
       volume = {947},
       number = {2},
          eid = {47},
        pages = {47},
          doi = {10.3847/1538-4357/acbfa9},
archivePrefix = {arXiv},
       eprint = {2212.04438},
 primaryClass = {astro-ph.CO},
       adsurl = {https://ui.adsabs.harvard.edu/abs/2023ApJ...947...47K}
}

@article{Smoot1991,
       author = {{Smoot}, G.~F. and others},
        title = "{First results of the COBE satellite measurement of the anisotropy of the cosmic microwave background radiation}",
      journal = {Advances in Space Research},
         year = 1991,
        month = jan,
       volume = {11},
       number = {2},
        pages = {193-205},
          doi = {10.1016/0273-1177(91)90490-B},
       adsurl = {https://ui.adsabs.harvard.edu/abs/1991AdSpR..11b.193S}
}

@article{Bennett_2003,
    author = "Bennett, C. L. and others",
    collaboration = "WMAP",
    title = "{First year Wilkinson Microwave Anisotropy Probe (WMAP) observations: Preliminary maps and basic results}",
    eprint = "astro-ph/0302207",
    archivePrefix = "arXiv",
    doi = "10.1086/377253",
    journal = {\apjs},
    volume = "148",
    pages = "1--27",
    year = "2003"
}

@article{Bennett_2013,
       author = {{Bennett}, C.~L. and {Larson}, D. and {Weiland}, J.~L. and {Jarosik}, N. and {Hinshaw}, G. and {Odegard}, N. and {Smith}, K.~M. and {Hill}, R.~S. and {Gold}, B. and {Halpern}, M. and {Komatsu}, E. and {Nolta}, M.~R. and {Page}, L. and {Spergel}, D.~N. and {Wollack}, E. and {Dunkley}, J. and {Kogut}, A. and {Limon}, M. and {Meyer}, S.~S. and {Tucker}, G.~S. and {Wright}, E.~L.},
        title = "{Nine-year Wilkinson Microwave Anisotropy Probe (WMAP) Observations: Final Maps and Results}",
      journal = {\apjs},
         year = 2013,
        month = oct,
       volume = {208},
       number = {2},
          eid = {20},
        pages = {20},
          doi = {10.1088/0067-0049/208/2/20},
archivePrefix = {arXiv},
       eprint = {1212.5225},
 primaryClass = {astro-ph.CO},
       adsurl = {https://ui.adsabs.harvard.edu/abs/2013ApJS..208...20B}
}

@article{plk13overview,
    author = "Ade, P. A. R. and others",
    collaboration = "Planck",
    title = "{Planck 2013 results. I. Overview of products and scientific results}",
    eprint = "1303.5062",
    archivePrefix = "arXiv",
    primaryClass = "astro-ph.CO",
    reportNumber = "CERN-PH-TH-2013-115",
    doi = "10.1051/0004-6361/201321529",
    journal = {\aap},
    volume = "571",
    pages = "A1",
    year = "2014"
}

@article{plk18compsep,
    author = "Akrami, Y. and others",
    collaboration = "Planck",
    title = "{Planck 2018 results. IV. Diffuse component separation}",
    eprint = "1807.06208",
    archivePrefix = "arXiv",
    primaryClass = "astro-ph.CO",
    doi = "10.1051/0004-6361/201833881",
    journal = {\aap},
    volume = "641",
    pages = "A4",
    year = "2020"
}

@article{plk2020npipe,
    author = "Akrami, Y. and others",
    collaboration = "Planck",
    title = "{$Planck$ intermediate results. LVII. Joint Planck LFI and HFI data processing}",
    eprint = "2007.04997",
    archivePrefix = "arXiv",
    primaryClass = "astro-ph.CO",
    doi = "10.1051/0004-6361/202038073",
    journal = {\aap},
    volume = "643",
    pages = "A42",
    year = "2020"
}

@ARTICLE{sevem2008,
       author = {{Leach}, S.~M. and others},
        title = "{Component separation methods for the PLANCK mission}",
      journal = {\aap},
         year = 2008,
        month = nov,
       volume = {491},
       number = {2},
        pages = {597-615},
          doi = {10.1051/0004-6361:200810116},
archivePrefix = {arXiv},
       eprint = {0805.0269},
 primaryClass = {astro-ph},
       adsurl = {https://ui.adsabs.harvard.edu/abs/2008A&A...491..597L}
}

@ARTICLE{sevem2012,
       author = {{Fern{\'a}ndez-Cobos}, R. and {Vielva}, P. and {Barreiro}, R.~B. and {Mart{\'\i}nez-Gonz{\'a}lez}, E.},
        title = "{Multiresolution internal template cleaning: an application to the Wilkinson Microwave Anisotropy Probe 7-yr polarization data}",
      journal = {\mnras},
         year = 2012,
        month = mar,
       volume = {420},
       number = {3},
        pages = {2162-2169},
          doi = {10.1111/j.1365-2966.2011.20182.x},
archivePrefix = {arXiv},
       eprint = {1106.2016},
 primaryClass = {astro-ph.CO},
       adsurl = {https://ui.adsabs.harvard.edu/abs/2012MNRAS.420.2162F}
}

\begin{appendices}

\counterwithin{figure}{section}
\counterwithin{table}{section}

\begin{figure}
\centering
\includegraphics[width=0.48\textwidth]{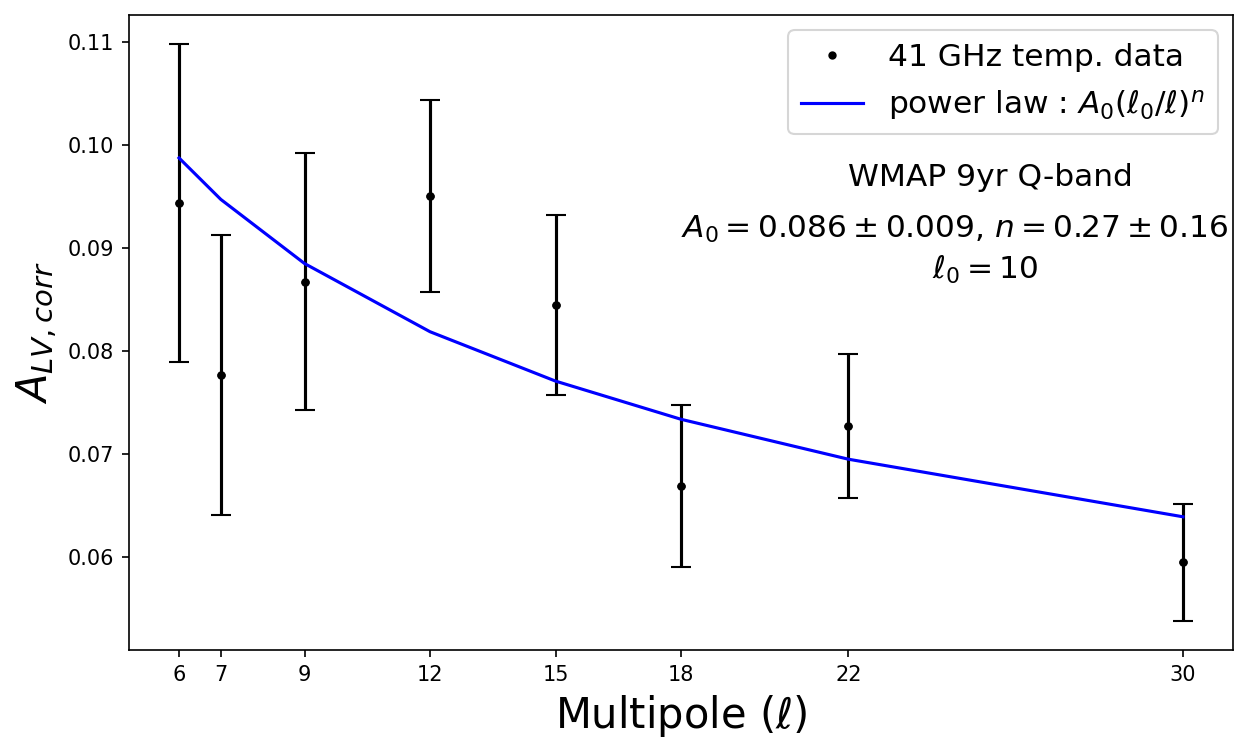}
~
\includegraphics[width=0.48\textwidth]{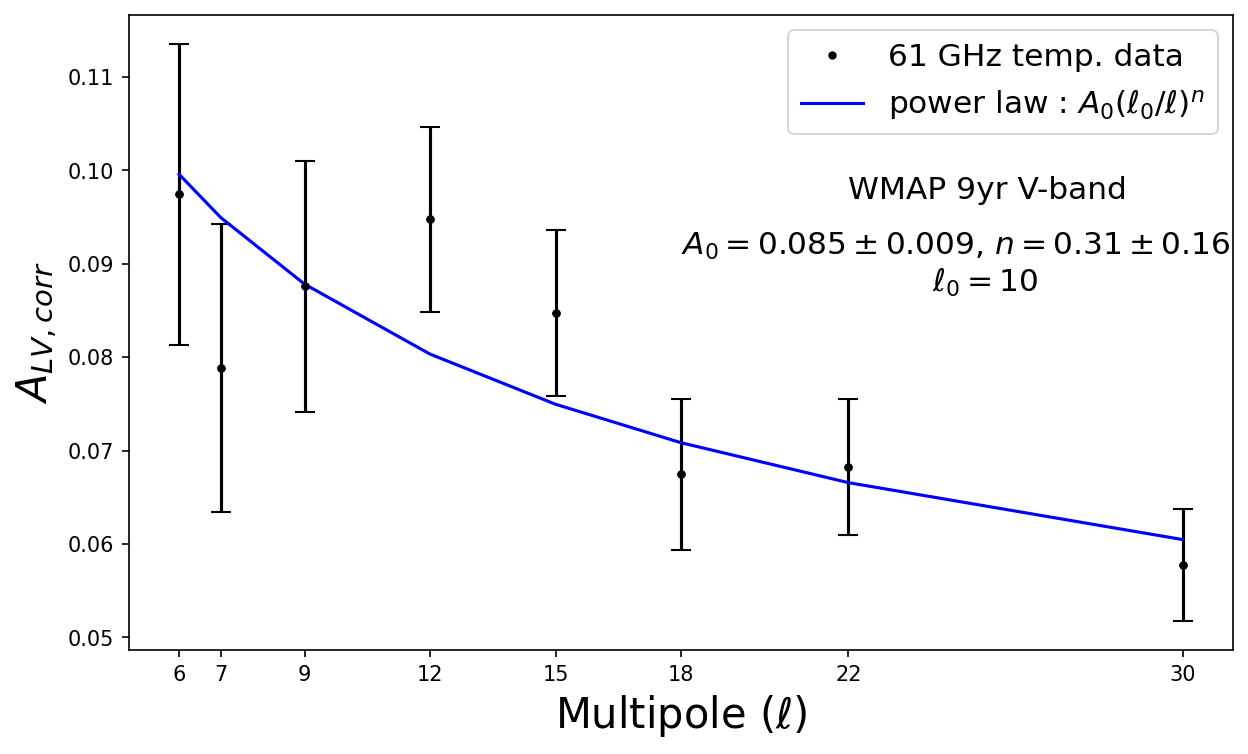}
~
\includegraphics[width=0.48\textwidth]{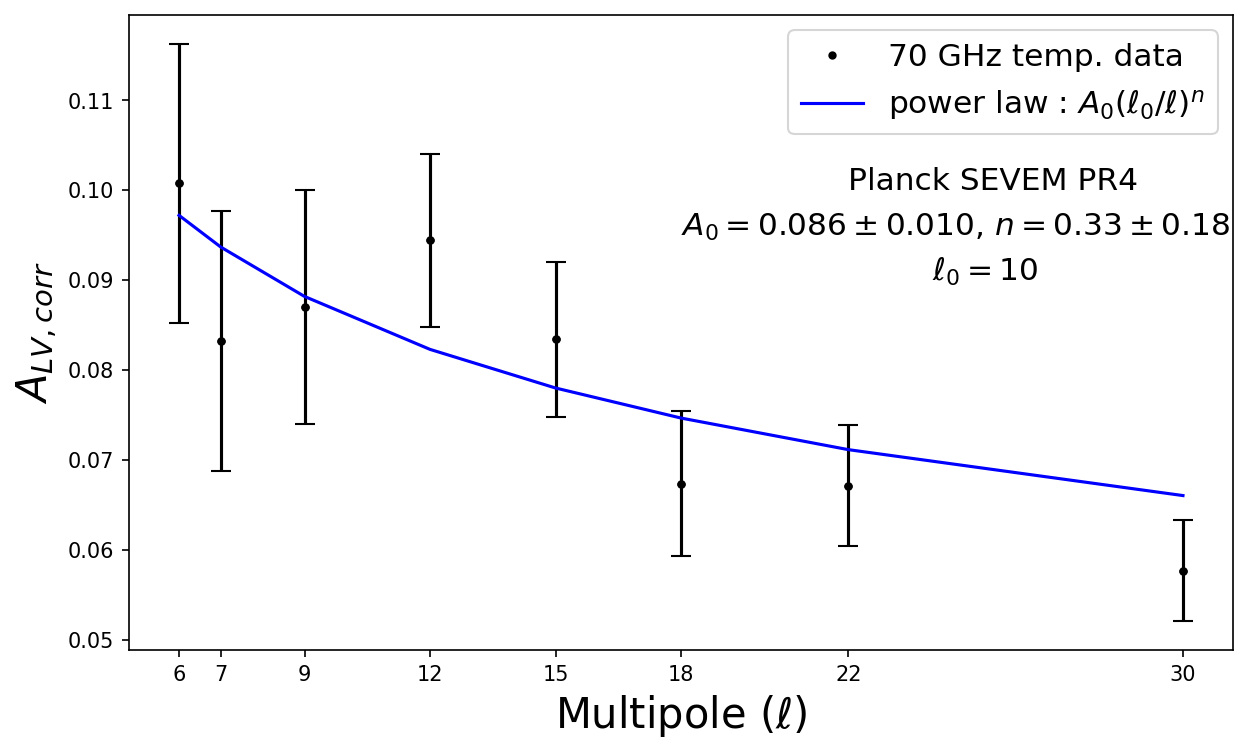}
~
\includegraphics[width=0.48\textwidth]{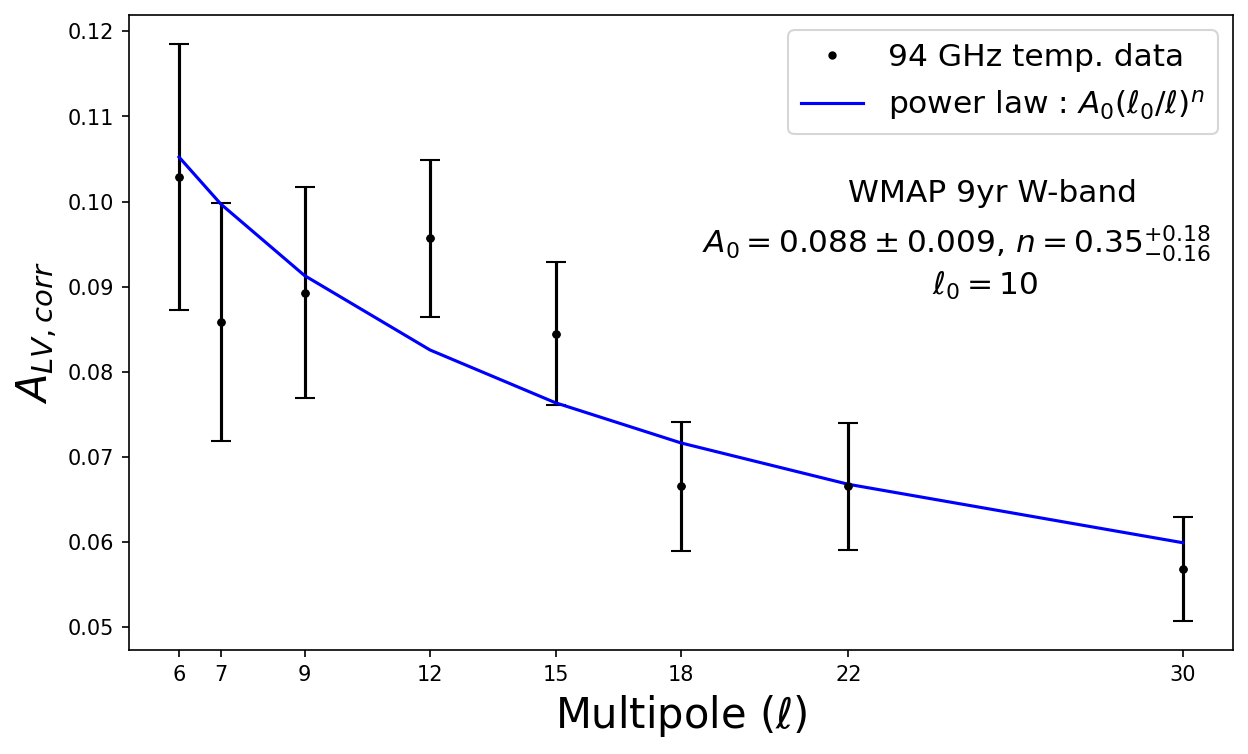}
~
\includegraphics[width=0.48\textwidth]{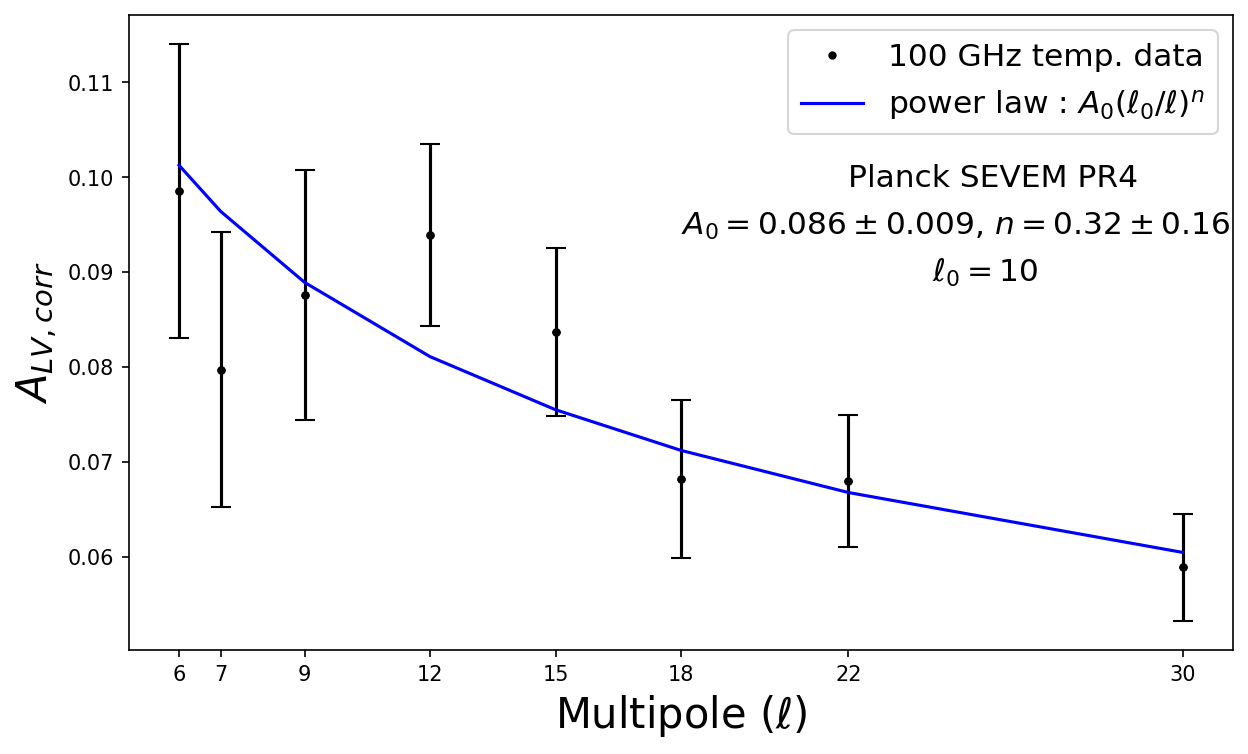}
~
\includegraphics[width=0.48\textwidth]{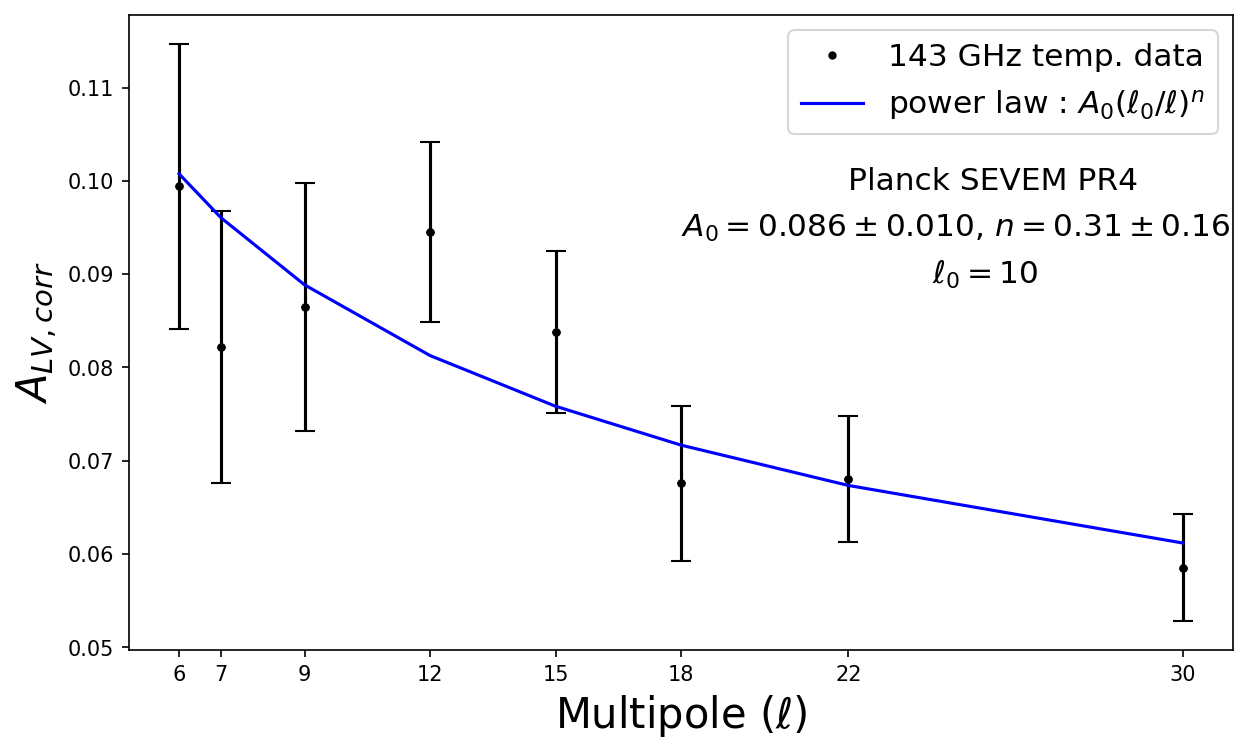}
~
\includegraphics[width=0.48\textwidth]{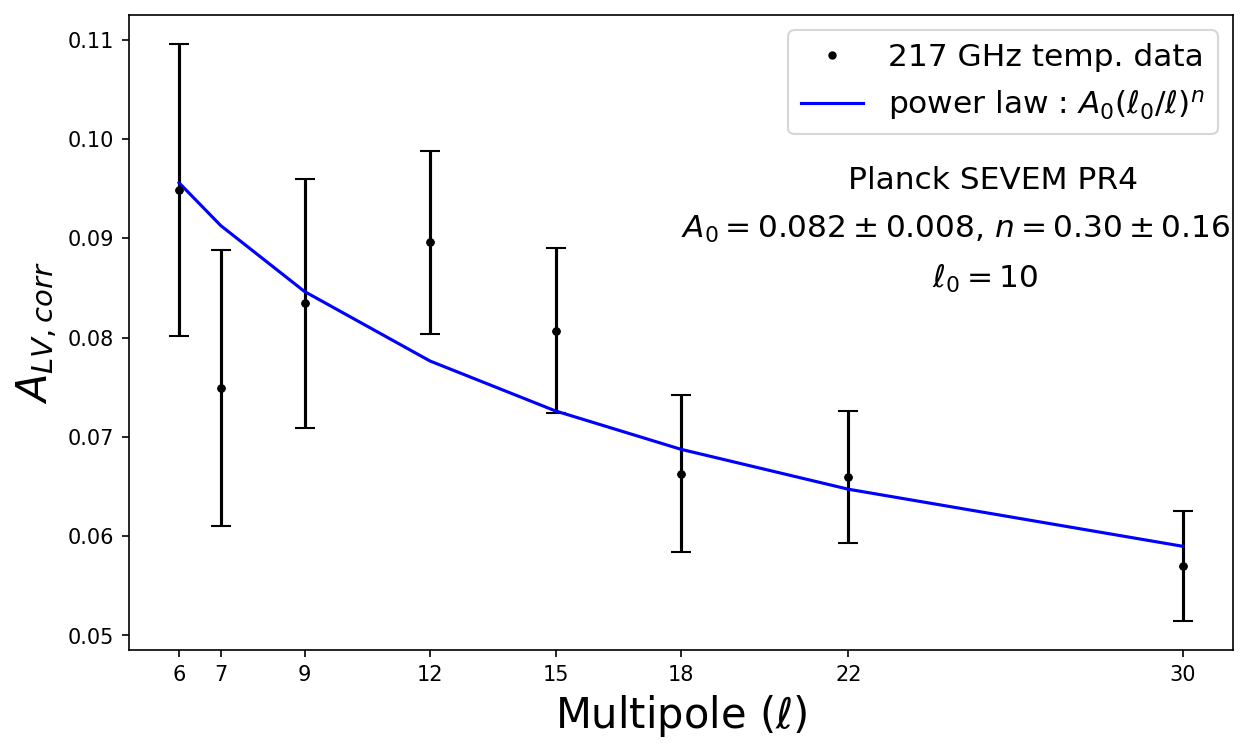}

\caption{Corrected dipole amplitudes, $A_{\rm LV,corr}\equiv A(\ell)$ (per Eq.~{\ref{eq:alv-bias-corr}}) from  LVE maps corresponding to various frequency-specific cleaned CMB temperature maps as a function of multipole `$\ell$' are fitted with a power-law model following Eq.~(\ref{eq:pl}) using MCMC method.}
\label{fig:apdx:power_law_par}
\end{figure}

\section{Power-law fitting of individual frequency-specific CMB maps using MCMC}
\label{apdx:pl-fit}

In the present analysis, we consider Eq.~(\ref{eq:pl}) as our theoretical model for testing scale dependence of hemispherical power asymmetry. For fitting the power-law, we used \texttt{Cobaya} package which does Bayesian inference via MCMC sampling of the underlying parameter distributions. MCMC doesn’t just find a single best fit value of the parameters, but rather maps out their full posteriors returning mean values and corresponding errors ($1\sigma$, $2\sigma$, etc., confidence levels as desired). In this method, parameter estimation is done by minimizing the $\chi^2$ function that is dependent on the model parameters which simply measures the ``distance'' between data and our chosen model. In our case, it is defined as, 
\begin{equation}
    \chi^2(A_{0}, n) = -2\ln({\mathcal{L}}) = \sum_\ell \left[\frac{A_{\rm obs}(\ell) - A_0(\ell_0/\ell)^n}{\sigma_\ell}\right]^2
    \label{eq:apdx:chi2}
\end{equation}
where `$A_{0}$' and `$n$' are the power-law fit parameters that we are interested in, $A_{\rm obs}(\ell)=A_{\rm LV,corr}$ are the bias-corrected LVM dipole amplitudes from data at different `$\ell$', and $\sigma_\ell$ are the corresponding (symmetrized) errors (that are obtained by averaging upper and lower errors derived from normalized LVMs corresponding to isotropic noisy CMB realizations). Further, $\mathcal{L}(A_0,n)$ is the associated likelihood function.
The likelihood is related to $\chi^2$ function in a way that the minimization of $\chi^2$ results in a maximization of $\mathcal{L}$.

We set the priors as $A_{0} = [0, 0.2]$ and $n = [-2,2]$ that are sufficiently wide. Also we choose the pivot scale to be $\ell_0=10$. The initial segment of an MCMC chain, during which the sampler is still locating the high probability regions of the parameter space, is referred to as the burn-in phase. The first 1,000 samples of the MCMC chain were discarded as burn-in, ensuring that only samples from the stationary posterior distribution (out of $\sim10,000$ total accepted steps) were used in computing the posteriors, mean and $1\sigma$ error bars.

The results of power-law fit to observed dipole amplitudes from the ``seven'' frequency-specific foreground cleaned CMB maps studied in the present work are shown in Fig.~[\ref{fig:apdx:power_law_par}]. The data derived LVM dipole amplitudes are shown using black dots along with $1\sigma$ error bars. The effect of cosmic variance is visible in the low multipole range, where the error bars are large. The curve in \emph{blue} represents the variation in dipole amplitude with multipole assuming the power-law with best fit values of the amplitude `$A_{0}$' and index `$n$' obtained following MCMC fitting. The scale dependence is similar across all frequency channels.

\end{appendices}

%
%

\end{document}